**Optimization and Validation of the DESIGNER dMRI preprocessing pipeline in white matter aging**

Jenny Chen[1], Benjamin Ades-Aron[1], Hong-Hsi Lee[2], Subah Mehrin[1], Michelle Pang[3], Dmitry S. Novikov[1], Jelle Veraart[1], Els Fieremans[1]

[1]Center for Biomedical Imaging (CBI), Center for Advanced Imaging Innovation and Research (CAI[2]R), Department of Radiology, New York University School of Medicine, New York, NY, USA
[2]Athinoula A. Martinos Center for Biomedical Imaging, Massachusetts General Hospital, Boston, MA, USA
[3]John A. Burns School of Medicine, University of Hawai'i at Manoa, Honolulu, HI, USA

Send correspondence to:
Jenny Chen
Center for Biomedical Imaging
Department of Radiology
New York University School of Medicine
E-mail: Jenny.Chen@nyulangone.org
660 First Avenue
New York, 10016

**ABSTRACT**

Various diffusion MRI (dMRI) preprocessing pipelines are currently available to yield more accurate diffusion parameters. Here, we evaluated accuracy and robustness of the optimized Diffusion parameter EStImation with Gibbs and NoisE Removal (DESIGNER) pipeline in a large clinical dMRI dataset and using ground-truth phantoms. DESIGNER, a preprocessing pipeline targeting various imaging artifacts in diffusion MRI data, has been modified to improve denoising and target Gibbs ringing for partial Fourier acquisitions. We compared the revised DESIGNER (Dv2) (including denoising, Gibbs removal, correction for motion, EPI distortion, and eddy currents) against the original DESIGNER (Dv1) pipeline, minimal preprocessing (including correction for motion, EPI distortion, and eddy currents only), and no preprocessing on a large clinical dMRI dataset of 524 control subjects with ages between 25 and 75 years old. We evaluated the effect of specific processing steps on age correlations in white matter with DTI and DKI metrics. We also evaluated the added effect of minimal Gaussian smoothing to deal with noise and to reduce outliers in parameter maps compared to DESIGNER-Dv2's noise removal method. Moreover, DESIGNER-Dv2's updated noise and Gibbs removal methods were assessed using a ground truth dMRI phantom to evaluate accuracy. Results show age correlation in white matter with DTI and DKI metrics were affected by the preprocessing pipeline, causing systematic differences in absolute parameter values and loss or gain of statistical significance. Both in clinical dMRI and ground-truth phantoms, DESIGNER-Dv2 pipeline resulted in the smallest number of outlier voxels and improved accuracy in DTI and DKI metrics as noise was reduced and Gibbs removal was improved. Thus, DESIGNER-Dv2 provides more accurate and robust DTI and DKI parameter maps by targeting common artifacts present in dMRI data acquired in clinical settings, as compared to no preprocessing or minimal preprocessing.

## 1 INTRODUCTION

Diffusion MRI (dMRI) has the ability to access in-vivo tissue microstructure (Novikov et al., 2019) noninvasively with sensitivity to healthy development, aging, and disease pathology. Quantitative parameters extracted from dMRI are potential biomarkers for early diagnosis and disease-specific treatment and/or prevention planning (Andica et al., 2020; Billiet et al., 2015; Moura et al., 2019). However, raw dMRI in clinic and clinical research applications are unavoidably contaminated by thermal noise and various image artifacts caused by pulse sequence, gradient hardware, acquisition strategy, and patient movement (Jones & Cercignani, 2010; Le Bihan et al., 2006). All these artifacts propagate into diffusion parameter maps, yielding unreliable results to investigators and clinicians.

Common artifacts present in raw dMRI include distortions caused by eddy currents when applying strong gradients in different directions, motion (Le Bihan et al., 2006), and low signal-to-noise ratio (SNR) due to diffusion weighting and the relatively long echo time (Dietrich et al., 2001; Jones & Cercignani, 2010). Although single-shot echo planar imaging (EPI) is fast and the standard acquisition strategy, it causes EPI distortion resulting from local magnetic field inhomogeneities (Le Bihan et al., 2006). In addition, due to finite sampling in k-space, dMRI suffers from Gibbs ringing in regions with high contrast boundaries (Kellner et al., 2016; Perrone et al., 2015; Veraart, Fieremans, et al., 2016). In particular, clinical dMRI is routinely done in partial Fourier acquisition mode, leading to additional ringing after reconstruction via zero filling. All these different sources of artifacts contribute to unphysical dMRI parameter estimates and can appear as unfeasible values (Dubkov & Malakhov, 1976; Henriques, Correia, et al., 2021; Kuder et al., 2012; Olson et al., 2018; Veraart, Fieremans, et al., 2016) in parametric maps.

Various dMRI preprocessing pipelines have been implemented to minimize some or all of these artifacts (Ades-Aron et al., 2018; Cieslak et al., 2021; Cui et al., 2013; Glasser et al., 2013; Irfanoglu et al., 2017; Maximov et al., 2019; Tournier et al., 2019), aiming to improve the accuracy and precision of outcome parameter maps. As numerous artifacts exist, dMRI pre-processing pipelines can range in complexity depending on study protocol and software availability. For example, a minimal preprocessing pipeline from the Human Connectome Project (HCP) (Glasser et al., 2013) corrects for EPI distortion, eddy currents, and motion, while more complex pipelines such as Diffusion parameter EStImation with Gibbs and NoisE Removal (DESIGNER) (Ades-Aron et al., 2018) additionally targets removal of common noise and Gibbs ringing artifacts. With many pipelines and correction techniques to choose from, MRtrix3 (Tournier et al., 2019), TORTOISE (Irfanoglu et al., 2017), and DIPY (Garyfallidis et al., 2014) are amongst many other software tools that make a range of previously proposed artifact correction methods available to the community and allow users to build a customized pipeline specific to their needs. It has been observed that the inclusion of these additional preprocessing steps improves the performance of widely used preprocessing steps such as motion and eddy current distortion correction (Cieslak et al., 2024). However, As observed by Maximov et al. (Maximov et al., 2019), pipeline variations not only affect diffusion metrics, they can possibly result in conflicting findings in the same study. There is a need to evaluate these correction steps to decide which techniques are optimal so we can confidently set up a preprocessing pipeline for future large-scale processing and interpret the study results.

In this study, we revisited the originally proposed DESIGNER (Diffusion parameter EStImation with Gibbs and NoisE Removal) (Ades-Aron et al., 2018) pipeline as a targeted artifact correction pipeline for processing typical dMRI as acquired in a clinical setting. The DESIGNER pipeline is updated to include improved denoising and Gibbs ringing removal methods suitable for partial Fourier acquisitions. We will call this updated DESIGNER, DESIGNER-Dv2. To evaluate and demonstrate the impact of varying preprocessing on statistical outcomes and results, we applied DESIGNER-Dv2, minimal preprocessing pipeline, and no preprocessing to a large clinical dMRI dataset (N=524). Then, to assess the accuracy and precision of these new techniques, we created ground truths and induced Gibbs ringing or added noise for evaluation.

The outline of this paper is as follows: We first described the new correction options, and then compared the effect of DESIGNER-Dv2 pipeline against a minimal preprocessing pipeline and no preprocessing on conventional dMRI parameter estimation. We compared pipelines of different complexities by looking at age correlation after preprocessing with the pipelines in white matter of controls, an extensively studied topic (Beck et al., 2021; Kodiweera et al., 2016; Maximov et al., 2019; Ouyang et al., 2021; Schilling et al., 2022; Taha et al., 2022; Toschi et al., 2020; Yeatman et al., 2014). In addition, we evaluated the accuracy and precision of the modified noise removal and Gibbs correction steps through ground truth comparison. We also assessed the added effect of Gaussian smoothing in the minimal preprocessing pipeline in comparison to applying noise removal. For this study, we focused on evaluating the artifact correction steps only, hence excluding any steps beyond this, such as outlier detection (Andersson et al., 2016; Chang et al., 2012), smoothing, and tensor estimation methods (Collier et al., 2015; Tristán-Vega et al., 2012; Veraart et al., 2013). This study introduces a more robust and accurate DESIGNER-Dv2 pipeline as an optimized preprocessing pipeline for conventional multi-shell dMRI as acquired in clinical (research) settings.

## 3 METHODS

### *3.1 Clinical Data*

We retrospectively studied control subjects (N=524, 363 females, 161 males) with ages ranging from 25- to 75-year-old. They were selected out of 5399 subjects who came in for routine clinical brain MRI on a Magnetom Prisma 3T (N=262) or Skyra 3T (N=262). Control subjects were identified by clinical indication of dizziness or headache without MRI abnormalities and without history of neurological disease upon chart reviews (Pang et al., 2022).

dMRI was acquired with the following protocol: 5 *b* = 0 images, 1 *b* = 0 image was acquired with reverse phase-encoding direction for EPI distortion correction (Andersson et al., 2003; Smith et al., 2004), b = 250 s/mm² – 4 directions, b = 1000 s/mm² – 20 directions, b = 2000 s/mm² – 60 directions, TE = 70ms (N=142) or 95ms (N=382), TR = 3.7s, 50 slices, resolution = 1.7x1.7x3mm³, 6/8 partial Fourier, generalized auto-calibrating partially parallel acquisition.

These clinical dMRIs included acquiring clinical diffusion tensor imaging (DTI) series (b = 0, 1000 s/mm²) alongside a research high b-value series (b = 0, 250, 2000 s/mm²) with the same imaging parameters (including TE) in both series. The high b-value series allows us to be sensitive to non-gaussian diffusion (Jensen & Helpern, 2010) and extract diffusion kurtosis imaging (DKI) metrics. However, high b-values require strong gradients and longer TE (applied to all b-values), which leads to increased distortion, lower signal-to-noise ratio (SNR), and longer scan time (Dietrich et al., 2001; Jones & Cercignani, 2010). Also, as a two series protocol, there may be intensity variations between the two series. Furthermore, this dMRI acquisition scheme came with several potential artifacts including Gibbs ringing from partial Fourier and EPI distortion, making this an ideal dataset to test preprocessing pipelines.

### *3.2 Simulation I- HCP noise phantom*

HCP phantom was generated as described by Ades-Aron et al. (Ades-Aron et al., 2018). The HCP phantom was then modified by projecting spherical harmonics along directions matching the clinical data described above so it can mimic the clinical data. Next, it was downsampled to 2.5x2.5x2.5mm³ (72 slices) and used as ground truth. Then, complex Gaussian noise was added using the equation $S_m = \sqrt{(S_r + \sigma\epsilon_1)^2 + (\sigma\epsilon_2)^2}$ where $S_m$ is the simulated Rician-distributed magnitude MRI signal, $S_r$ is the real-valued signal, $\sigma$ is the noise level, and $\epsilon_1$ and $\epsilon_2$ are independent normal variables with zero mean and unit variance, to create 50 sets of HCP phantom with SNR of 10, 15, 20, 25, 30, and 60. Note, without loss of generality, the imaginary signal component was set to be zero. Varying the SNRs showed the denoising effect on varying degrees of noise.

### *3.3 Simulation II- Shepp-Logan phantom for Gibbs ringing evaluation*

Numerical simulation of the 2D Shepp-Logan phantom, a mathematical model of a brain made up of ellipses with varying size and signal intensities (Jain, 1989; Shepp & Logan, 1974), was used to assess Gibbs ringing removal methods. As the HCP phantom was originally created from dMRI datasets acquired with partial Fourier, it could not be used to evaluate partial Fourier-induced Gibbs ringing removal methods because it had remaining Gibbs ringing from partial Fourier. Instead, Gibbs ringing from k-space truncation and 6/8 partial Fourier in the horizontal direction was introduced to Shepp-Logan phantom.

### *3.4 Experiments*

#### *3.4.1 Preprocessing pipelines*
Figure 1 shows the flow diagram of the main pipelines employed on clinical data in this study. Below lists all pipelines applied on clinical data:

1. The minimal preprocessing pipeline targets EPI distortion (Andersson et al., 2003), eddy current (Andersson & Sotiropoulos, 2016), and motion correction (Andersson et al., 2016). We call this Eddy/topup+Motion (E+M) pipeline. Since the clinical dataset included two series, E+M also includes b0 normalization, which does voxelwise rescaling by taking the ratio of smoothed (Gaussian kernel (mm) with 3 standard deviations) $b$=0 images from each series to rescale all diffusion-weighted images. This pipeline was based on the routinely used HCP diffusion preprocessing pipeline (Glasser et al., 2013).
2. Original DESIGNER (Dv1) is a previously validated (Ades-Aron et al., 2018) pipeline involving the same corrections as the E+M pipeline defined above. It additionally deals with thermal noise, Gibb ringing, and Rician signal biases using local patch MPPCA denoising (Cordero-Grande et al., 2019; Veraart, Novikov, et al., 2016), subvoxel-shifts (SuShi) Gibbs removal (Kellner et al., 2016), and Rician bias correction (Koay & Basser, 2006), respectively.
3. DESIGNER (Dv2) is an updated version of DESIGNER-Dv1. Recent technical developments promote a more efficient denoising and wider applicable Gibbs removal method, as outlined by the following updates:
    a. Adaptive patch MPPCA denoising with eigenvalue shrinkage (Gavish & Donoho, 2017) in DESIGNER-Dv2 creates patches of voxels from the same tissue type to maximize signal redundancy during denoising and, as such, the denoising performance. The adaptive patch selection uses a bilateral approach to select 100 voxels with the most similar underlying tissue content. For the central voxel in the patch, we choose voxels with both small Euclidean distance to this voxel and voxels with the smallest difference in intensity (over all diffusion directions).
    b. Removal of Partial-fourier induced Gibbs ringing (RPG) (Lee et al., 2021) was developed to target ringing from both symmetric and asymmetric k-space truncation. Partial Fourier acquisition is common in clinical (research) settings and leads to both types of ringing. While SuShi (Kellner et al., 2016) Gibbs removal works on ringing from symmetric k-space truncation, asymmetric k-space truncation ringing pattern remains. This has led to the development of RPG (Lee et al., 2021), which additionally removes ringing from asymmetric k-space by resampling the image so SuShi Gibbs removal targets this remaining ringing on the resampled image. RPG works best on data with 6/8 and 7/8 partial Fourier.
4. To compare the DESIGNER-Dv2 pipeline with one that simply uses smoothing to deal with noise, we applied E+M pipeline with minimal Gaussian smoothing (full width at half maximum (FWHM) of 1.2*voxel size) to the clinical data.
5. To study the effect of adaptive patch MPPCA denoising with eigenvalue shrinkage (including Rician bias correction) separately from RPG, DESIGNER-Dv2 as described above, but without Gibbs correction was also applied to clinical data.
6. To study the effect of RPG separately from denoising and Rician bias correction, DESIGNER-Dv2 without denoising and Rician bias correction was applied to the clinical data.

E+M, DESIGNER-Dv1, and DESIGNER-Dv2 pipelines were applied using options within DESIGNER available for installation at https://github.com/NYU-DiffusionMRI/DESIGNER-v2.

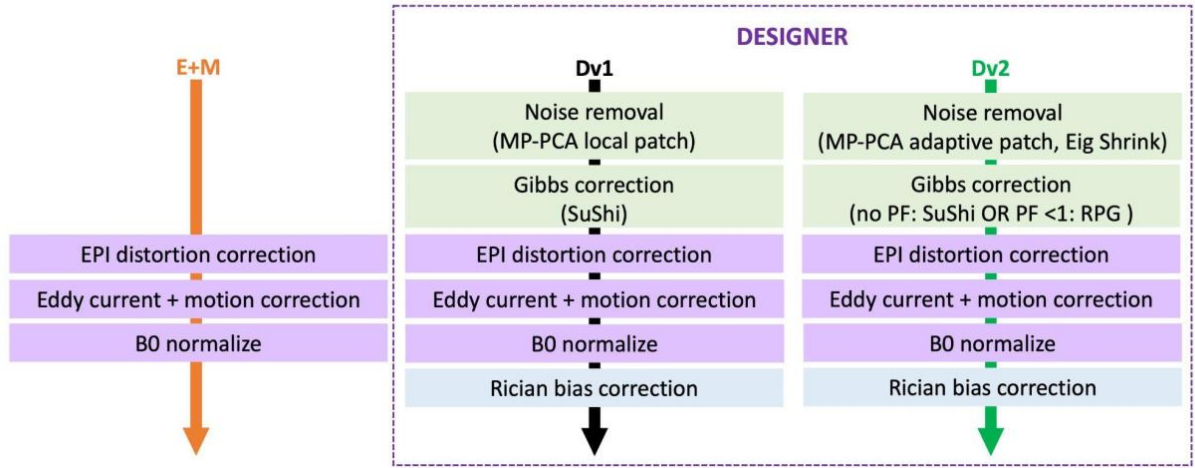

**Figure 1.** Step-by-step flow diagram for E+M, original DESIGNER (Dv1), and current DESIGNER (Dv2). Eig Shrink: eigenvalue shrinkage, PF: Partial Fourier, SuShi: Subvoxel-Shifts, RPG: Removal of Partial Fourier induced Gibbs.

*3.4.2 Clinical dMRI experiments*

Using the clinical dMRI scans, we tested how applying pipelines with different complexities affect parameter outcomes using a minimal preprocessing pipeline versus a more advanced preprocessing pipeline involving noise removal and Gibbs ringing removal. DTI (MD – mean diffusivity (µm$^2$/ms), RD – radial diffusivity (µm$^2$/ms), AD – axial diffusivity (µm$^2$/ms), and FA – fractional anisotropy) and DKI (MK – mean kurtosis, RK – radial kurtosis, AK – axial kurtosis) parameter maps from clinical data were extracted using weighted linear least squares fit (Veraart et al., 2013) after preprocessing the DWIs (diffusion-weighted images) in three different ways: (1) no artifact or noise removal, (2) E+M, and (3) DESIGNER-Dv2. Then, they were compared visually and quantitatively. Clinical data was also preprocessed with DESIGNER-Dv1 and compared to DESIGNER-Dv2.

Then, to look at the effect of smoothing to deal with noise, clinical dMRI scans were also preprocessed with E+M pipeline in addition to CSF excluded Gaussian smoothing with FWHM of 1.2*voxel size (E+M+Smooth). Parameter maps were then compared for those without preprocessing, E+M, E+M+Smooth, DESIGNER-Dv1, and DESIGNER-Dv2 pipelines applied (figure 2).

Outlier voxels across all DTI/DKI parametric maps were defined based on their physically possible lower and upper bounds: diffusivity D (MD, AD, RD) > 0, fractional anisotropy 0 < FA < 1, and kurtosis (Dubkov & Malakhov, 1976) K (MK, AK, RK) > -2. Outliers are values outside of these bounds (the upper bound of D < 3 is rarely violated, so it was not included here). We calculated average outlier percentage across parameter maps over each ROI (described below) to quantify remaining outlier voxels after preprocessing with each pipeline.

For quantitative measures, to exclude boundary voxels affected by partial volume effect, JHU white matter atlas regions of interest (ROIs) (Hua et al., 2008) were each clipped in atlas space to exclude 5% of the lowest FA. They were then warped to each participant's FA map by first applying a linear registration using FSL's FLIRT (FMRIB's Linear Image Registration Tool) (Jenkinson et al., 2002; Jenkinson & Smith, 2001) which was used to initialize the nonlinear warp using FNIRT (Andersson et al., 2010). For studying the effect of the pipeline on age correlations, we extracted the median from each parameter map for ROIs of the posterior limb of internal capsule (PLIC), splenium (SCC), genu of corpus callosum (GCC), and anterior corona radiata (ACR). The median was selected to reduce the effect of outliers in the ROIs, which varies between pipelines (figure 3). These regional metrics were then plotted with respect to age. Note that our data contains images acquired with either TE = 70ms or TE = 95ms. To avoid confounding effects due to mixing TEs and scanners, age correlations were plotted for data acquired by the same TE and scanner (Prisma, Skyra).

A quadratic fit was plotted to show age association as adjusted $R^2$ values using quadratic fit is greater when compared to adjusted $R^2$ using linear fit for all age associations (supplementary table S1). Bonferroni adjusted P-values (multiply p-value by 28 (7 parameters x 4 ROIs)), concavity, and age when extremum of age dependence is reached (age of decline) were collected from each parameter-ROI pairing to compare the effect pipelines have on age association. Significance level was set to Bonferroni adjusted P-values < 0.05. Since the age range is 25 to 75 years-old, the quadratic fit for that range may not include an extremum. In that case, instead of indicating concavity and age when extremum of age dependence is reached, we indicated if DTI/DKI parameters are monotonically increasing or decreasing.

Additionally, we considered the effects of denoising and Gibbs ringing removal separately on the clinical data, by preprocessing with pipelines (5) and (6) described above. Percent difference from E+M derived metrics were quantified for each scan in SCC, GCC, ACR, and PLIC ROIs (mean of ROI after omitting outliers).

### 3.4.3 Simulation experiments

To evaluate the updated denoising method in DESIGNER-Dv2, various denoising methods were applied to the HCP noise phantoms. Phantoms were only preprocessed with noise removal and/or Rician bias correction as we only introduced complex Gaussian noise to the ground truth and we were only interested in comparing the denoising methods here. They were preprocessed using (1) E+M's (no denoising), (2) Dv1's (local patch denoising followed by Rician bias correction), and (3) Dv2's (adaptive patch denoising with eigenvalue shrinkage followed by Rician bias correction) denoising method. Noisy phantoms were also preprocessed with (4) adaptive patch denoising with eigenvalue shrinkage (without Rician bias correction) and (5) adaptive patch denoising without eigenvalue shrinkage followed by Rician bias correction to compare effect of Rician bias correction and eigenvalue shrinkage separately.

For each parameter, maps were computed as the median over 50 noise realizations for a given SNR level. We took the median rather than the mean to avoid maps resulting in only outliers for low SNR phantoms. Denoising methods were evaluated by calculating, for each denoising method, a median percentage error map against their respective ground truth parameter maps. Similar to clinical data, JHU white matter atlas ROIs were warped to HCP phantom's ground truth FA map, but then merged into an overall WM ROI. We then extracted and plotted the median over the WM ROI from the median percentage error map to compare percent error after applying the different denoising methods. Additionally, average outlier percentage across parameter maps over WM ROI was calculated to quantify remaining outlier voxels, defined as $D < 0$, $K < -2$ or $FA > 1$, after each denoising method.

To evaluate smoothing to deal with noise, CSF excluded Gaussian smoothing with full width at half maximum (FWHM) of 1.2*voxel size) was applied on a set of HCP noise phantoms (SNR 10, 15, 20, 25, 30, 60). Then, we extracted median percent error in SCC from parameter maps yielded from: (1) no noise removal (2) Gaussian smoothing, (3) adaptive patch denoising with eigenvalue shrinkage followed by Rician bias correction, and (4) adaptive patch denoising without eigenvalue shrinkage followed by Rician bias correction.

The Gibbs Shepp-Logan phantom was preprocessed with E+M's (no Gibbs removal), Dv1's (Subvoxel-shifts Gibbs removal), or Dv2's (RPG) Gibbs correction method to evaluate the updated Gibbs removal method in DESIGNER-Dv2. Only Gibbs removal was used on the phantom since we are only evaluating the Gibbs removal methods here and no other image artifact was introduced to the phantom. Gibbs ringing removal methods were assessed by generating mean percentage error against ground truth phantom over four manually drawn ROIs targeting Gibbs ringing artifacts. Smoothing effects from applying weighting filters and simulation with full-Fourier acquisition have been evaluated previously by (Lee et al., 2021), and thus not explored here.

# 4 RESULTS

## *4.1 Clinical MRI*

*4.1.1 Effect of preprocessing pipelines on age association in white matter of clinical data*

Figure 2 shows FA, MD, and MK maps of a 60-year-old female, which appeared least noisy after preprocessing with DESIGNER-Dv2 as compared to no preprocessing, E+M, E+M+Smooth, or DESIGNER-Dv1. In the SCC/CSF boundary, the Gibbs artifact in SCC/CSF boundary (bright band in FA, dark band in MD, and black voxels in MK) was greatly reduced after preprocessing with RPG from DESIGNER-Dv2. In addition, the MK maps show improvement as the number of negative outliers or black voxels drastically decreased with E+M+Smooth, DESIGNER-Dv1, and DESIGNER-Dv2 when compared to no preprocessing and E+M.

The effect of preprocessing on the presence of outliers in parametric maps of all 524 clinical dMRI is shown in figure 3, plotting the percent outliers in PLIC, SCC, GCC, and ACR after no preprocessing, E+M, E+M+Smooth, DESIGNER-Dv1, and DESIGNER-Dv2 preprocessing. SCC generally has the most outliers, followed by PLIC, GCC, and finally ACR. Results from Tukey's HSD test for multiple comparisons show percent outliers in each of the four ROIs were significantly different between all pipelines (p-value < 0.05) except between DESIGNER-Dv1 and DESIGNER-Dv2 in PLIC, E+M+Smooth and DESIGNER-Dv1 in PLIC, SCC, and GCC, E+M+Smooth and DESIGNER-Dv2 in GCC, and between E+M, E+M+Smooth, DESIGNER-Dv1, and DESIGNER-Dv2 in ACR. Employing more advanced pipelines decreased the percentage of outliers in ROIs. The systematic reduction in outliers after E+M+Smooth, DESIGNER-Dv1 and DESIGNER-Dv2 preprocessing is in line with what we see in figure 2.

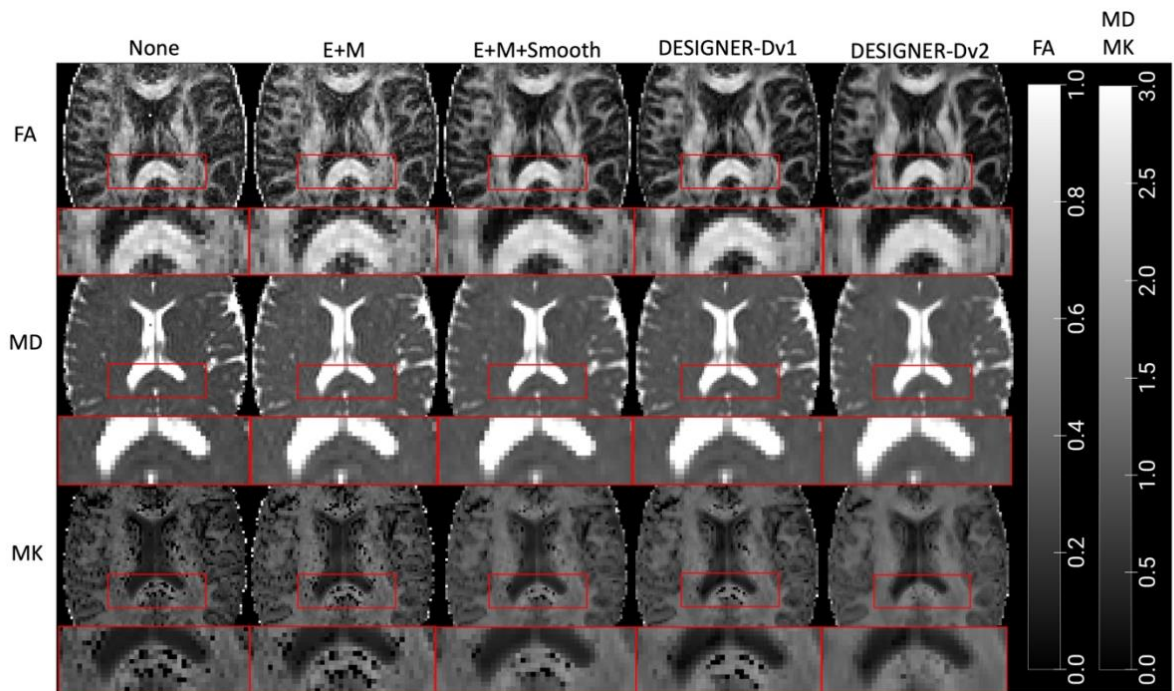

**Figure 2.** Parameter maps (FA, MD, MK) of a healthy 60-year-old female based on preprocessing of DWI in 5 different ways: (1) no preprocessing, (2) E+M, (3) E+M+Smooth (CSF excluded Gaussian smoothing with FWHM=1.2*voxel size), (4) DESIGNER-Dv1, and (5) DESIGNER-Dv2 pipeline. DESIGNER-Dv2 appears to result in parameter maps with the least noise and outliers, and greatly reduced Gibbs ringing in the splenium/corpus callosum boundary (bright band in FA, dark band in MD, and black voxels in MK).

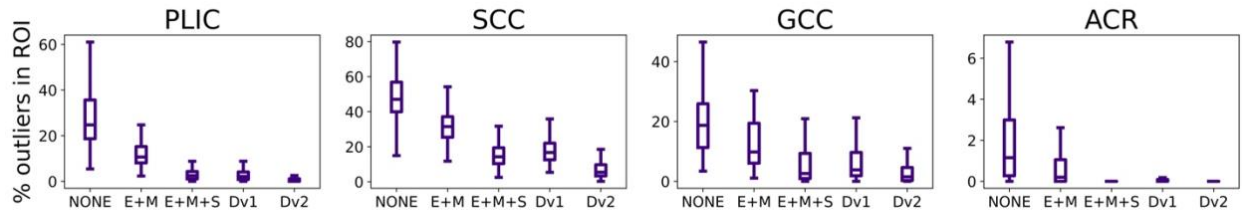

**Figure 3.** Boxplots of percent outliers (100*number of outliers in ROI/number of voxels in ROI) in each ROI of the dMRI parameter maps of 524 subjects based on no preprocessing, E+M, E+M+Smooth, DESIGNER-Dv1, and DESIGNER-Dv2 preprocessing pipeline. Tukey's HSD test for multiple comparisons show percent outliers in each of the four ROIs were significantly different between all pipelines (p-value < 0.05) except between DESIGNER-Dv1 and DESIGNER-Dv2 in PLIC, E+M+Smooth and DESIGNER-Dv1 in PLIC, SCC, and GCC, E+M+Smooth and DESIGNER-Dv2 in GCC, and between all pipelines except no preprocessing in ACR.

Figure 4 shows age correlations with DTI/DKI parameters of 142 control subjects (TE = 70ms, Prisma) derived from DESIGNER-Dv2 and E+M and no preprocessing (none) in ACR, PLIC, GCC, and SCC. For all ROIs and parameters, the observed age-dependencies were similar in shape for the curves obtained with different preprocessing pipelines. However, as also observed in histograms found in supplementary figure S3, the type of preprocessing resulted in some systematic differences in absolute parameter values: compared to no preprocessing and E+M, when preprocessed with DESIGNER-Dv2 pipeline, (1) FA and AK were systematically lower while (2) RD, MK and RK were systematically higher. Preprocessing pipelines also had varying impact in different metrics. For example, Figure 4 reveals that preprocessing pipelines did not affect MD as much as other metrics in all ROIs. Similar results are observed in figures S7 and S9 for data acquired with TE = 95ms.

The strength of the observed age-associations also depended on the preprocessing pipeline. In table S1 adjusted $R^2$ in ACR and GCC consistently increases from no preprocessing to DESIGNER-Dv2 for MD, RD, RK, MK and FA. Correlations either increased or decreased in strength with different pipelines (table S2). For example, for MK and RK in ACR and GCC, from no preprocessing to DESIGNER-Dv2, correlations increased as more advanced preprocessing pipeline was applied. On the other hand, correlations decreased in strength for the AD in all ROIs from no preprocessing to DESIGNER-Dv2. Age correlation results for the remaining data (TE = 95ms for Prisma and Skyra) are shown in supplementary section (tables S3-6). Overall, p-values and adjusted $R^2$s from TE = 95ms data agreed with results from TE = 70ms data (tables S1-2).

Supplementary figures S2, S4, S6, S8, S10 compares DESIGNER-Dv1 and DESIGNER-Dv2. Overall, age association adjusted $R^2$ and p-values (tables S1-2) were similar between the DESIGNER pipelines. Some systematic differences in absolute parameter values were observed in figure S2 (Prisma, TE = 70ms), S8 (Prisma, TE = 95ms), and S10 (Skyra, TE = 95ms): compared to DESIGNER-Dv1, preprocessing with DESIGNER-Dv2 had systematically (1) higher RD and MK and (2) lower FA. Additionally, figure S6 compares coefficient of variation defined by standard

deviation/mean of ROI and ROI coefficient of variation from DESIGNER-Dv2 was consistently lower than DESIGNER-Dv1.

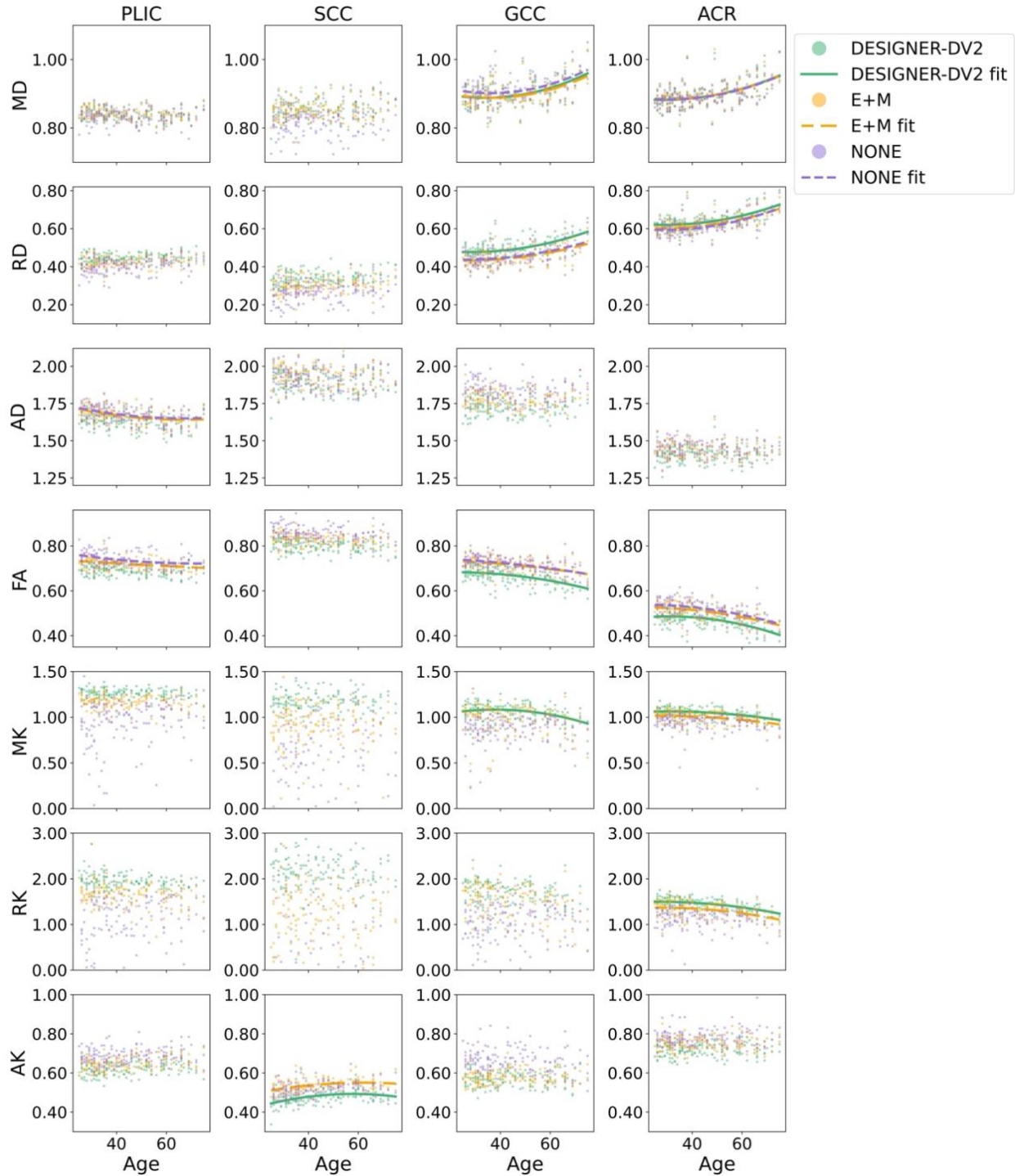

**Figure 4.** Age correlation with DTI and DKI parameters in white matter ROIs (median) from DESIGNER-Dv2 pipeline, E+M, or not preprocessing of 142 healthy subjects (Prisma, TE = 70ms). Quadratic fits were plotted for statistically significant correlations

with adjusted $R^2 > 0.1$ only. MK and RK plots are zoomed in so some outlier datapoints (figure 3) are not shown.

Table 1 summarizes concavity, age when extremum of age dependence was reached, and if parameters were monotonically increasing or decreasing from 25 to 75 years-old for statistically significant age associations with adjusted $R^2 > 0.1$. Based on the type of preprocessing pipeline, there was disagreement on whether the quadratic fit was in the concave portion of the quadratic model or monotonically increasing or decreasing for FA in PLIC and ACR.

**Table 1.** Age at peak (in years) from DTI and DKI age correlation quadratic fits in white matter regions using no preprocessing, E+M, and DESIGNER-Dv2 preprocessing pipeline. Red cells indicate concaved down (negative curvature) quadratic fit, and blue cells indicate concaved up (positive curvature) quadratic fit. (-) = monotonically decreasing from ages 25-75, (+) = monotonically increasing from ages 25-75. Only statistically significant age associations with adjusted $R^2 > 0.1$ are shown.

|      |             | MD    | RD    | AD    | FA    | MK    | RK    | AK    |
|------|-------------|-------|-------|-------|-------|-------|-------|-------|
| PLIC | None        |       |       | 66.97 | 72.98 |       |       |       |
|      | E+M         |       |       | 67.01 | (-)   |       |       |       |
|      | DESIGNER-Dv2|       |       |       |       |       |       |       |
| SCC  | None        |       |       |       |       |       |       |       |
|      | E+M         |       |       |       |       |       |       | 62.20 |
|      | DESIGNER-Dv2|       |       |       |       |       |       | 57.80 |
| GCC  | None        | 36.20 | 25.43 |       | (-)   |       |       |       |
|      | E+M         | 36.34 | 27.14 |       | (-)   |       |       |       |
|      | DESIGNER-Dv2| 33.96 | 27.80 |       | (-)   | 37.94 |       |       |
| ACR  | None        | 31.20 | 28.34 |       | (-)   |       |       |       |
|      | E+M         | 30.05 | 26.80 |       | (-)   |       | 25.46 | 27.96 |
|      | DESIGNER-Dv2| 27.63 | 28.59 |       | 30.57 |       | 29.59 | 26.52 |

*4.1.2 Effect from updated denoising versus Gibbs correction*

The effect of DESIGNER-Dv2's adaptive patch denoising with eigenvalue shrinkage (and Rician bias correction) and RPG in clinical data are quantified separately with respect to E+M in figure 5 as box plots. Results from one-way ANOVA showed that the effect of adaptive patch denoising with eigenvalue shrinkage versus RPG was significantly different for all metrics in each ROI (p-value < 0.05) except MD in ACR and AD in PLIC and GCC. The boxplots reveal denoise and bias correction can have opposite effects from Gibbs correction (AK in all ROIs, FA in GCC). There was also generally a greater effect from adaptive patch denoising with eigenvalue shrinkage than from RPG. Additionally, DESIGNER-Dv2's denoise and Gibbs removal steps affected MK and RK in each ROI more than DTI metrics.

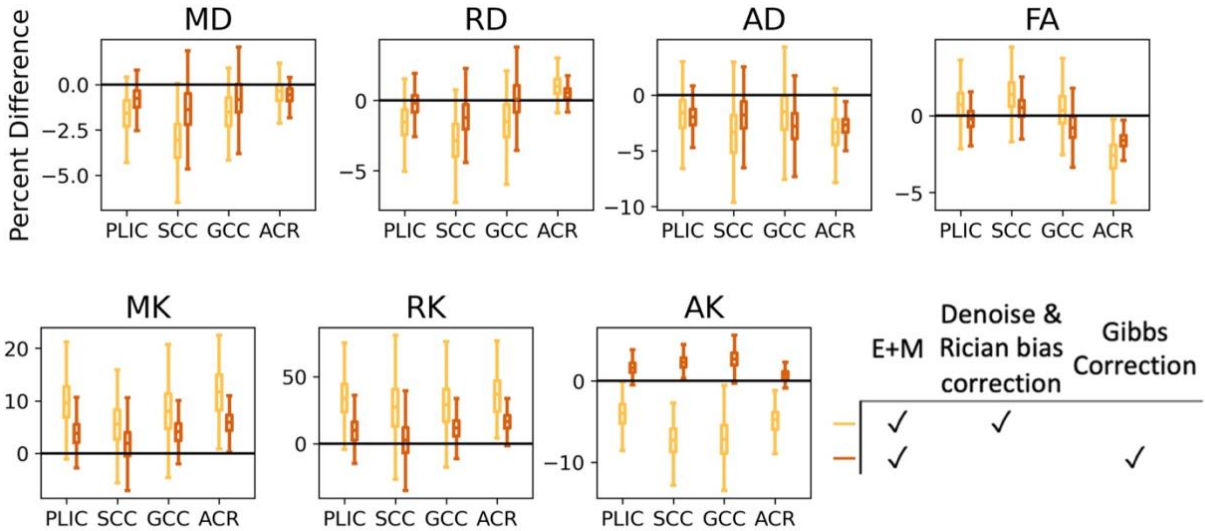

**Figure 5.** Boxplots of mean percent difference from E+M after denoising with adaptive patch with eigenvalue shrinkage (yellow) or Gibbs ringing removal with RPG (red) for 524 healthy subjects on ROI level (mean after omitting outliers). One-way ANOVA showed the effect of adaptive patch denoising with eigenvalue shrinkage versus RPG was significantly different for all metrics in each ROI (p-value < 0.05) except MD in ACR and AD in PLIC and GCC.

*4.1.3 Preprocessing time*

The preprocessing time for each clinical dataset did not differ much from pipeline to pipeline. The most time-consuming step was Eddy current and motion correction. On a 24 CPU core Ubuntu Linux server with 256 GB memory, E+M completed in approximately 45 minutes while noise and Gibbs removal from DESIGNER-Dv1 and DESIGNER-Dv2 took up an additional 5 and 7 minutes to complete respectively.

***4.2 Simulations***

*4.2.1 Assessment of denoising methods on HCP phantom*

Figure 6 shows FA, MD, and MK maps of HCP noise phantom (SNR=20) with varying denoising methods applied. We observed reduced outliers due to noise when processing with denoising methods from Dv1 and Dv2 versus no denoising (None). We also observed adaptive patch (Dv2) slightly improved denoising performance over local-patch (Dv1). Furthermore, eigenvalue shrinkage (Dv2) appeared to have an additional denoising effect, yielding less noise and outlier voxels.

Figure 7A-G plots median over the WM ROI from median percentage error maps. The plots show no denoising resulted in the most bias in all parameters except AK and denoising without Rician bias correction had the most bias in AK. Overall, denoising with an adaptive patch without eigenvalue shrinkage yielded the most accurate result. Interestingly, denoising with eigenvalue shrinkage generally had some bias in the opposite direction of the noise bias in MD, RD, AD, and FA maps. This bias from eigenvalue shrinkage was opposite to the effect of eigenvalue repulsion. For example, with eigenvalue shrinkage, FA and AD had negative bias and RD had positive bias. However, both figures 6 and 7H show kurtosis maps with eigenvalue shrinkage had the fewest outlier voxels. Outlier percentage in WM decreased from 0.27% to 0.10% when using adaptive patch followed by Rician bias correction versus adaptive patch with eigenvalue shrinkage followed by Rician bias correction respectively on noisy phantom with SNR=20. Eigenvalue shrinkage appeared to over-correct for noise (more bias with more noise) to give a smoothing effect on the parameter maps.

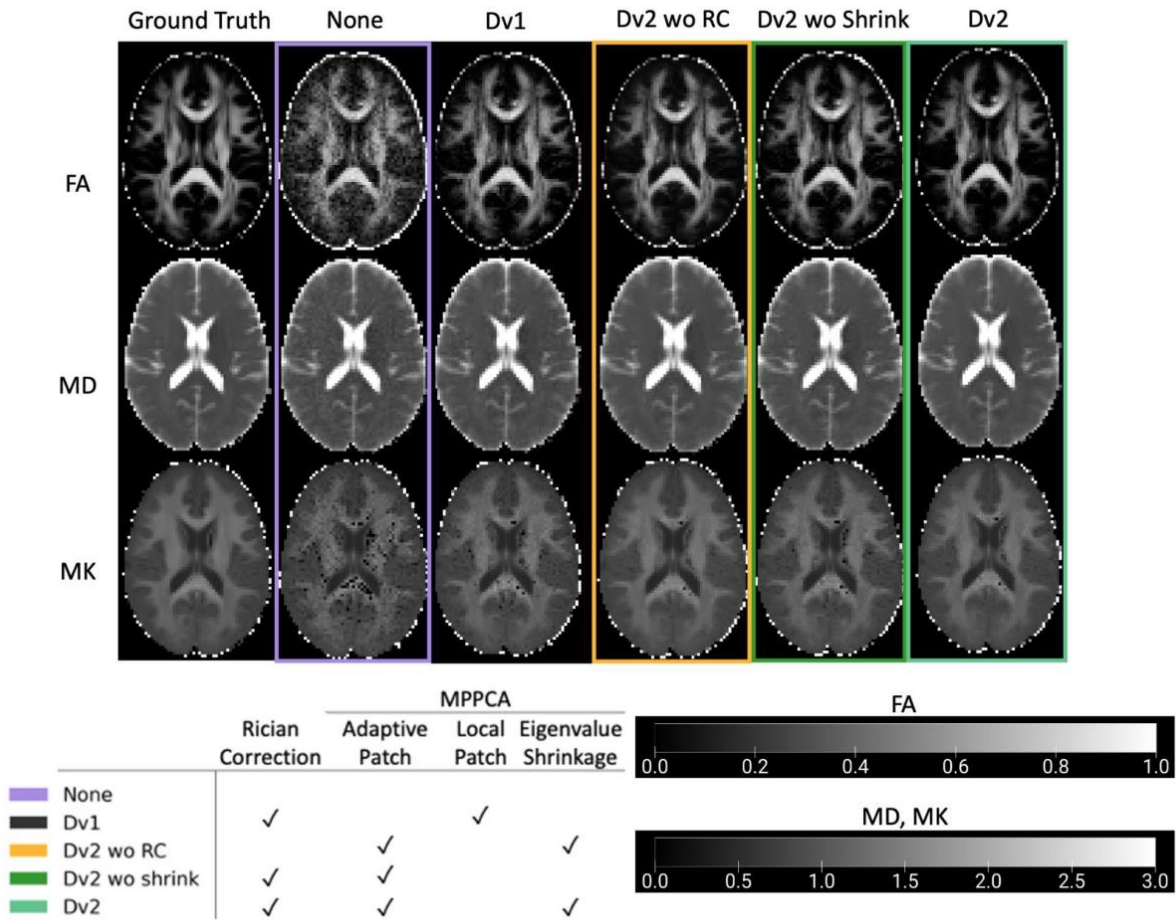

**Figure 6.** FA, MD, and MK maps for ground truth HCP phantom and noisy phantom (SNR 20) with varying denoising methods. Maps with adaptive patch denoising with eigenvalue shrinkage applied appeared to have the least black voxels.

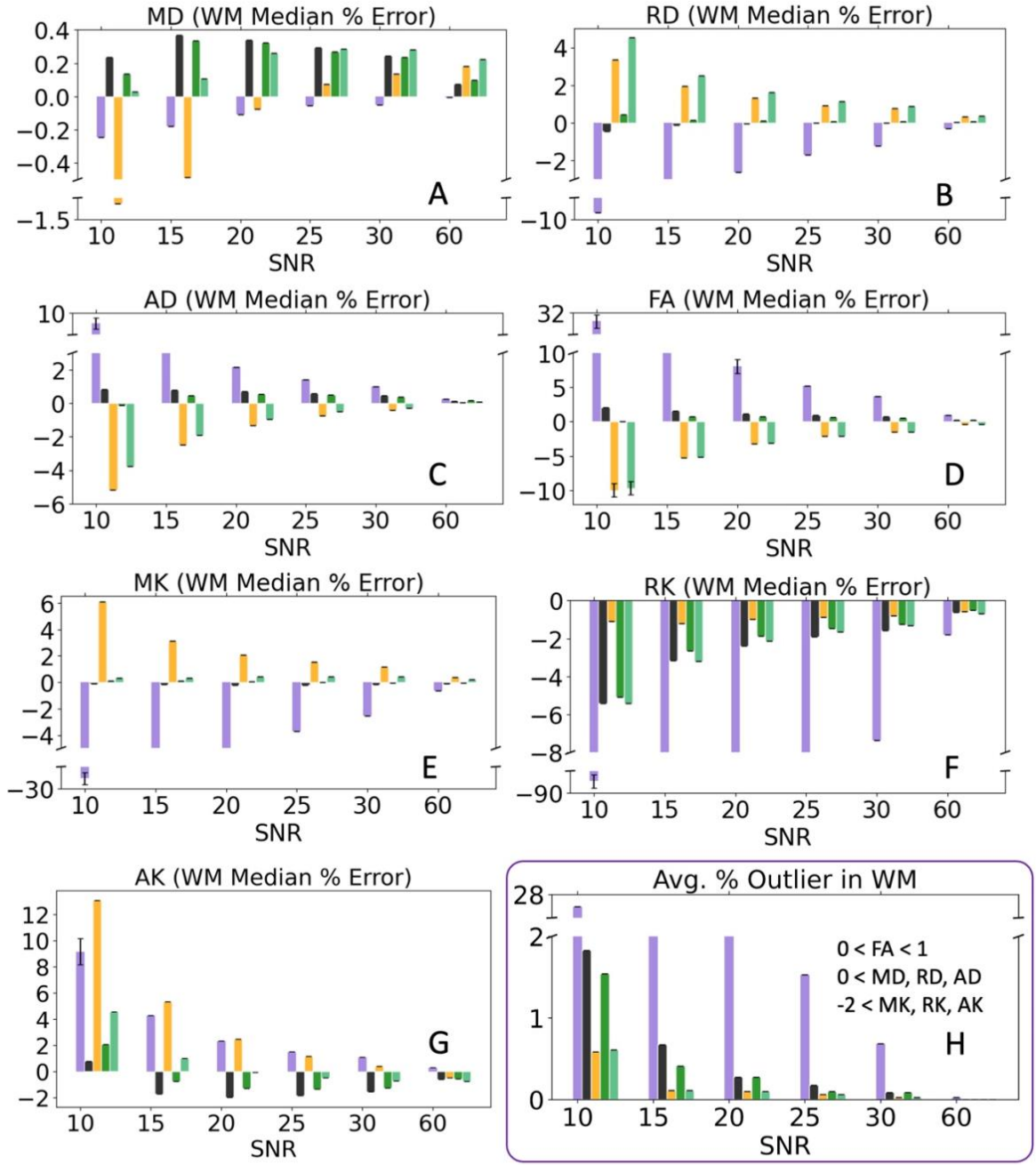

**Figure 7.** Ground truth evaluation of noise removal methods. A-G: Bar plots showing median and interquartile range (error bar scaled by 1/10) of DTI/DKI parameters median
16

percentage error in WM of 50 sets of phantoms (SNR from 10-60). H: Average percent outlier (WM) of 50 phantom iterations (SNR from 10-60). Note for SNR between 30 and 60, denoising became obsolete in MD (where median percent error is only within 1.5% across all SNR even without any correction) and AK.

Figure 8 compares median percent error from ground truth in SCC after smoothing and denoising a set of HCP noise phantoms. At low SNR of 10-20, smoothing improved accuracy in all metrics except AK. However, at SNR of 25-60, smoothing worsened accuracy in all metrics besides MK and RK. Except at low SNR of 10-20 in MD, adaptive patch denoising with eigenvalue shrinkage followed by Rician bias correction (Dv2) and/or adaptive patch denoising without eigenvalue shrinkage followed by Rician bias correction (Dv2 wo shrink) performed better than smoothing.

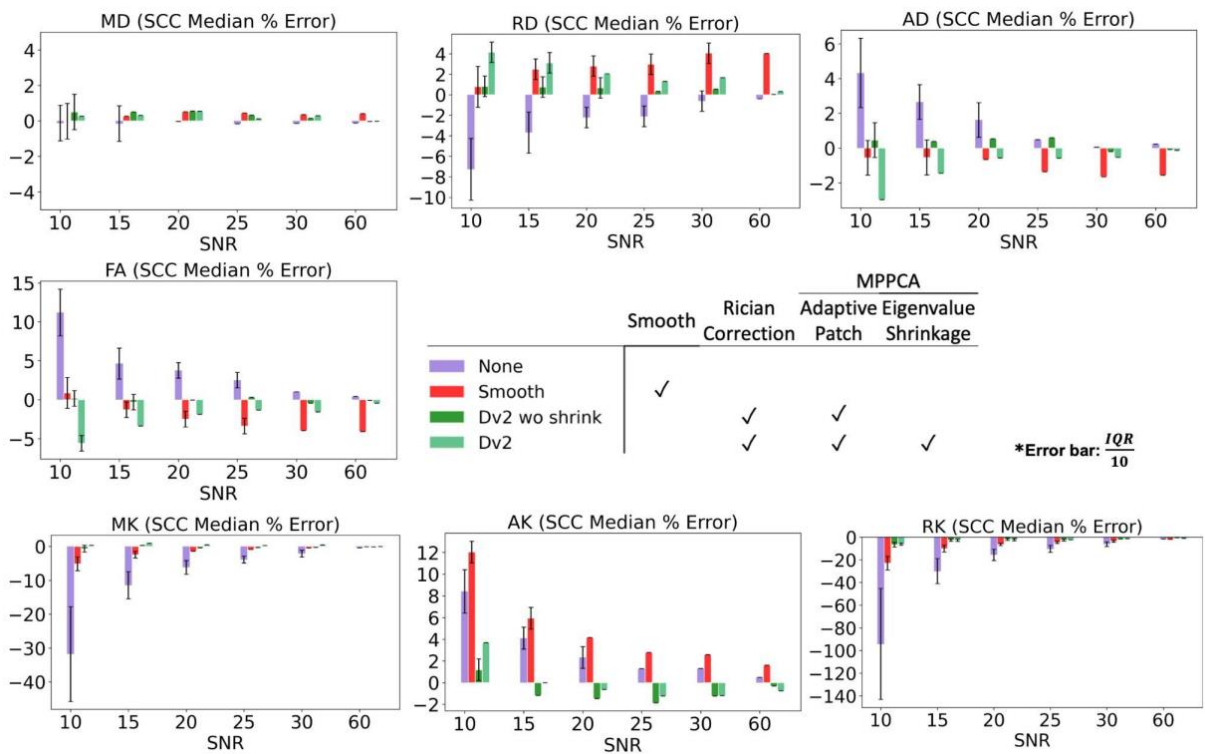

**Figure 8**. Ground truth evaluation of CSF excluded Gaussian smoothing and noise removal methods. Bar plots show median percentage error and interquartile range (error bar scaled by 1/10) of DTI/DKI parameters in SCC of a set of noise phantoms (SNR from 10-60).

*4.2.2 Assessment of Gibbs ringing correction on Shepp-Logan phantom*

Figure 9A shows DTI/DKI mean percentage error maps for no Gibbs correction and Gibbs correction using SuShi or RPG. The maps show using the SuShi Gibbs correction method removed some Gibbs ringing artifacts while using RPG removed additional ringing resulting from partial Fourier. Figure 9B shows mean percentage error in each manually drawn ROI targeting Gibbs ringing. Gibbs removal using RPG was the most accurate, followed by SuShi method, while not applying correction had the greatest percent error in all ROIs and parameters.

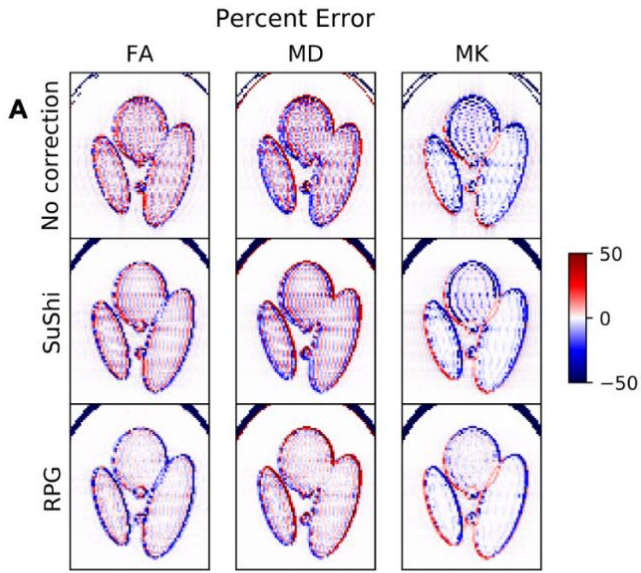
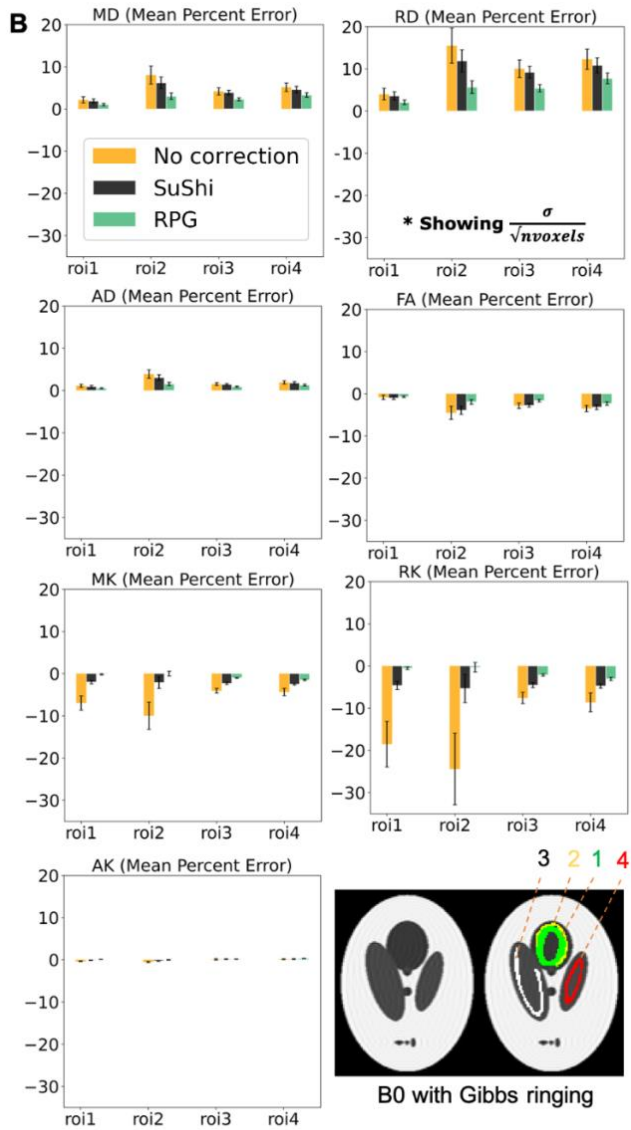

**Figure 9.** Evaluation of Gibbs removal. DTI/DKI mean percentage error compared between Shepp-Logan phantom with Gibbs ringing (simulated from k-space truncation and 6/8 PF in the horizontal direction) without correction, corrected with SuShi method, and corrected with RPG. A: FA, MD, and MK mean percentage error maps. B: Bar plots showing ROIs' mean percentage error and standard error in DTI/DKI parameter maps.

## 5 DISCUSSION

This study reveals improved accuracy and robustness of DTI/DKI voxelwise estimation and thus more reliable results when multi-shell dMRI acquisitions are preprocessed with DESIGNER-Dv2 pipeline compared to no preprocessing, E+M, and DESIGNER-Dv1 pipeline. Our age correlation study displays how using different preprocessing pipelines induces variability in parameter values. Visualizing parameter maps helped us understand the importance of denoising and Gibbs removal for improving DTI and DKI maps. Additionally, ground truth results also show optimized correction methods from DESIGNER-Dv2 yield more accurate and robust maps. The pipeline comparisons using clinical dMRI along with the simulated ground truth studies illustrate that using DESIGNER-Dv2 pipeline yields more reliable diffusion parameter values.

To visualize the effect of noise and Gibbs removal, we looked at parameter maps from a healthy 60-year-old female. We confirmed reduced noise, Gibbs ringing, and outlier voxels in parameter maps as we introduced denoising and Gibbs ringing removal (figure 2). Considering all subjects, we also quantified percent outliers in ROIs and saw significant decrease in outliers as preprocessing steps were added (figure 3). Agreeing with figure 2, in ROIs such as corpus callosum, affected by Gibbs ringing, outliers were further reduced when targeting partial Fourier induced Gibbs ringing in DESIGNER-Dv2, which was not included in DESIGNER-Dv1. While a previous study (Kornaropoulos et al., 2022) did not find improved outcome for denoising and Gibbs removal when performing analyses based on ROI-values, our results show that parametric maps from minimal processing (E+M) feature profoundly more outliers as compared to our DESIGNER pipelines due to denoising and Gibbs removal. These observations show we can expect improved voxelwise statistical results when appropriate correction strategies in dMRI data are used.

Age correlations in our clinical data reveals results varied based on the employed preprocessing pipeline. Age correlations (figure 4) revealed systematic differences between parameter values from varying pipelines in all ROIs for all metrics except MD. Additionally, it is known from aging patterns in past literature, that late-myelinating tracts are more vulnerable to neurodegeneration (Bartzokis, 2004; Benitez et al., 2014). So, we expect changes with age due to demyelination reflected most in radial metrics (RD, RK) and FA in late-myelinating tracts (Bartzokis, 2004; Benitez et al., 2014; Paydar et al., 2014). Here (table S1), as expected, adjusted $R^2$ from no preprocessing to DESIGNER-Dv2 is consistently increasing in the more vulnerable late-myelinating tracts (ACR and GCC) as compared to SCC and PLIC. Since myelin affects diffusion mostly in the radial direction, there is no clear biological effect due to demyelination in the axial metrics expected. Indeed, the low correlation in AD with no preprocessing is likely spurious as it further weakens with improved pipeline, potentially due to improved artifact correction including Gibbs. Thus, preprocessing pipeline affects parameter values and statistical results. These findings highlight the importance of preprocessing all data with the same optimal pipeline and to be wary of mixing pipelines in studies.

To differentiate the effect of noise removal and the effect of Gibbs removal on DTI and DKI parameters from clinical data, we used E+M as the baseline of comparison. The effect of DESIGNER-Dv2's noise removal and Gibbs removal are significantly different, and the corrections can have opposing changes as they are targeting different issues. Thus, for more reliable metrics, both denoising and Gibbs correction should be applied to address noise and Gibbs ringing. Additionally, DESIGNER-Dv2's denoising (adaptive patch denoising with eigenvalue shrinkage) and Rician bias correction generally contributes more to the change in resulting DTI and DKI ROI values than RPG does (figure 5). This is likely because Gibbs ringing appears as oscillations, so taking the mean percent difference over an ROI then cancels out the error. The systematic differences in absolute parameter values (Figure 4) between E+M and DESIGNER-Dv2 are mostly contributed from denoising and Rician bias correction.

Ground truth phantoms were used to evaluate the recent denoising method proposed in our DESIGNER-Dv2 pipeline: MPPCA adaptive patch denoising (compared to local patch) with eigenvalue shrinkage and Rician bias correction. The absolute median percent error of DTI/DKI parameters in white matter was much greater when no noise removal was applied on phantoms with SNR up to 30 (figure 7), implying denoising at SNR of about 30 and below improves accuracy greatly. For SNR between 30 and 60, denoising became less beneficial (but also not harmful). While noise phantoms show denoising was the largest contributing factor to increase accuracy for most parameters, Rician bias correction contributed most to bias in AK.

The insights from phantom data helps interpret the observed systematic differences in the clinical diffusion maps (figure 4) as being likely due to lowering noise floor in AK and reduction of eigenvalue repulsion from improved denoising in RD, FA, MK, and RK when using DESIGNER-Dv2 pipeline. In addition, adaptive patch denoising generally results in the lowest median percent error whereas eigenvalue shrinkage biases the results. On the other hand, adaptive patch denoising with eigenvalue shrinkage has the fewest outliers and the least noise in maps (figure 6 and 7). Thus, while adaptive patch denoising gives the most accurate median value after omitting outliers, we set adaptive patch denoising with eigenvalue shrinkage as the default denoising in DESIGNER-Dv2 as it removes the most noise and excludes the least voxels due to outliers.

Alternatively, smoothing can be applied to reduce noise, as demonstrated in figures 2 and 8. Smoothing appears to reduce noise in clinical dMRI maps (figure 2) but does not perform as well as applying denoising in DESIGNER-Dv2. Similarly in noise phantoms (figure 8), smoothing reduces noise at low SNR for certain parameters, but adaptive patch denoising and Rician bias correction performs better, and smoothing may actually be more harmful for SNR>20. Thus, smoothing is a reasonable option if denoising cannot be applied but will not yield optimal results, and ideally should be preceded by denoising.

DESIGNER-Dv2's modified Gibbs removal method, RPG, was also evaluated using ground truth phantoms. RPG targets ringing due to partial Fourier, particularly for 7/8 and 6/8 partial Fourier. Lee et al. have demonstrated, correcting for 5/8 partial Fourier will result in unreasonably smoothed maps (Lee et al., 2021). Mean percent error in manually drawn ROIs (targeting the ringing artifact) on Shepp-Logan phantom with 6/8 partial Fourier shows improved accuracy using RPG compared to SuShi (figure 9). This can be visually verified in percent error maps (figure 9A). Hence, for clinical data with similar partial Fourier acquisition, we verify that it would be valuable to use RPG to effectively remove Gibbs ringing and improve accuracy in parameter maps.

Further improvements to yield more accurate parameter maps may be to adjust the shrinkage when denoising to minimize bias, consider signal drift, and to deal with remaining black voxels. Signal drift has been found to cause a global decrease in signal intensity due to temporal scanner instability and can vary from scanner to scanner (Vos et al., 2017). Vos et al. investigated the effects of signal drift in dMRI data and proposed to fit a nonlinear model using interspersed $b$=0 images from the scan to interpolate and calculate correction factors for the images. This is a similar correction to DESIGNER-Dv2's B0 normalization which can be adjusted and carried over to correct for signal drift. Evaluating signal drift in our clinical data and incorporating this correction in our pipeline may yield more accurate parameter estimates. Although the DESIGNER-Dv2 pipeline greatly reduces the number of outlier voxels, for the diffusion acquisition protocol employed in our study, we still cannot avoid problematic black voxels in kurtosis maps just from preprocessing. The remaining black voxels may be eliminated in the tensor fitting step via a combination of outlier detection and correction, smoothing, and robust fitting (Henriques, Jespersen, et al., 2021).

This work focused on evaluating denoising and Gibbs removal as additional components in image preprocessing pipelines. We conclude that the addition of denoising and Gibbs ringing correction improves the performance of the preprocessing pipeline. Similar findings were recently reported by (Cieslak et al., 2024) albeit their study was limited to denoising only. In its essence, we demonstrate the impact of improving individual preprocessing steps to maximize the performance of the whole pipeline, without ranking or quantifying the relative contribution of a single preprocessing step on the overall performance. Hence, our results do not suggest that the use of denoising or Gibbs ringing removal lowers the importance of other image preprocessing steps such as motion correction. Indeed, for example, dMRI datasets in this study were initially preprocessed on servers with different versions of FSL (v5.0.8 versus v6.0.5), giving us inconsistent results and prompting us to reprocess all data with FSL v6.0.5 on the same server. Figure S12 shows the importance of using an updated eddy and motion correction version by updating FSL as resulting maps are suboptimal regardless of denoising and Gibbs removal.

Various recent pipelines (Cai et al., 2021; Cieslak et al., 2021; Cui et al., 2013; Irfanoglu et al., 2017), including DESIGNER (Dv1 and Dv2), are similar in that they target the correction of artifact-specific approaches and are agnostic to further post-processing.

Thus, highlighted advances, including RPG, might represent relevant development for any other pipeline. Moreover, DESIGNER-Dv2 is developed and validated on a large cohort of clinical data that represent the realistic challenges and variability of clinical radiology. Therefore, we believe that DESIGNER-Dv2 is well-equipped to bridge the gap between clinical and research applications of neuroimaging.

DESIGNER is available at https://github.com/NYU-DiffusionMRI/DESIGNER-v2 where instructions are available for using DESIGNER with Docker or for installation via pip. Dependencies for running the pipelines in this paper include MRtrix3 (https://www.mrtrix.org), FMRIB Software Library (FSL (v6.0.5 was used for this study), https://fsl.fmrib.ox.ac.uk/fsl/fslwiki/FSL), and Python. However, additional dependencies (not necessary for E+M, Dv1, and Dv2) include DIPY (https://dipy.org) for Patch2Self (Fadnavis et al., 2020) and ANTs (https://github.com/ANTsX/ANTs) for ANTs motion correction options. We recommend making sure FSL versions are updated when preprocessing data for studies as older versions of FSL may not perform as well and mixing FSL versions can give inconsistent results. One way to ensure consistent pipeline is used would be to run DESIGNER using Docker. DESIGNER comes with detailed documentation (https://nyu-diffusionmri.github.io/DESIGNER-v2) for each available preprocessing option, so users will only need a basic understanding of dMRI data.

## 6 CONCLUSION

The updated DESIGNER (Dv2) pipeline provides accurate and robust parameters for clinical research studies. It is designed to be used on dMRI scans acquired by standard multi-shell dMRI clinical protocol with common artifacts. DESIGNER-Dv2 is flexible enough to allow steps to be left out or adjusted to cater to a wider range of images. The standard pipeline consists of: a) MPPCA adaptive patch denoising with eigenvalue shrinkage, b) RPG (PF<1) or SuShi (Kellner et al., 2016) (no PF) correction for Gibbs, c) EPI distortion correction, d) Eddy current and motion correction, e) B0 normalization, and f) Rician bias correction. Our study shows the benefit of using the DESIGNER-Dv2 pipeline on diffusion MRI with common artifacts as no preprocessing or minimal preprocessing may result in many outliers and unreliable parameter values.

## DATA AND CODE AVAILABILITY

Clinical data is available upon request. DESIGNER is available at https://github.com/NYU-DiffusionMRI/DESIGNER-v2 as a preprocessing tool for diffusion MRI.

## AUTHOR CONTRIBUTIONS

JC: Analysis, Visualization, Data curation, and Writing – original draft. BA: Software, Methodology, and Writing – review & editing. HL: Software, Methodology, and Writing – review & editing. MP: Data curation, Writing – review & editing. DN: Methodology, Funding acquisition, Writing – review & editing. JV: Methodology, Supervision, Writing – review & editing. EF: Methodology, Supervision, Funding acquisition, Writing – review & editing.

**DECLARATION OF COMPETING INTERESTS**

JV, DSN and EF are co-inventors of patent US20180120404A1 (licensed) describing the denoising technology evaluated in this study. HHL, DSN, and EF are co-inventors of the degibbsing technology (patent pending) evaluated in this study. DSN, EF, and BA are shareholders at Microstructure Imaging, Inc.

**ETHICS STATEMENT**

This retrospective study was reviewed and approved by the Institutional Review Board (IRB) in the United States (IRB number: i14-01224). Waiver of consent was obtained.

**ACKNOWLEDGMENTS**

We thank Saurabh Maithani for assistance in collecting data and Durga Kullakanda for helping characterize clinical data.

This work was supported by the National Institute of Neurological Disorders and Stroke of the NIH under award R01 NS088040, National Institute of Biomedical Imaging and Bioengineering of the NIH under award number R01 EB027075, Irma T. Hirschl Foundation, Office of the Director of the NIH and National Institute of Dental & Craniofacial Research of the NIH under award number DP5 OD031854 and was performed at the Center of Advanced Imaging Innovation and Research (CAI2R, www.cai2r.net), and NIBIB Biomedical Technology Resource Center (NIH P41 EB017183).

**REFERENCES**

Ades-Aron, B., Veraart, J., Kochunov, P., McGuire, S., Sherman, P., Kellner, E., Novikov, D. S., & Fieremans, E. (2018). Evaluation of the accuracy and precision of the diffusion parameter EStImation with Gibbs and NoisE removal pipeline. *NeuroImage*, *183*, 532–543. https://doi.org/10.1016/j.neuroimage.2018.07.066

Andersson, J. L. R., Graham, M. S., Zsoldos, E., & Sotiropoulos, S. N. (2016). Incorporating outlier detection and replacement into a non-parametric framework for movement and distortion correction of diffusion MR images. *NeuroImage*, *141*, 556–572. https://doi.org/10.1016/j.neuroimage.2016.06.058

Andersson, J. L. R., Jenkinson, M., & Smith, S. (2010). *Non-linear registration, aka Spatial normalisation*.

Andersson, J. L. R., Skare, S., & Ashburner, J. (2003). How to correct susceptibility distortions in spin-echo echo-planar images: Application to diffusion tensor imaging. *NeuroImage*, *20*(2), 870–888. https://doi.org/10.1016/S1053-8119(03)00336-7

Andersson, J. L. R., & Sotiropoulos, S. N. (2016). An integrated approach to correction for off-resonance effects and subject movement in diffusion MR imaging. *NeuroImage*, *125*, 1063–1078. https://doi.org/10.1016/j.neuroimage.2015.10.019

Andica, C., Kamagata, K., Hatano, T., Saito, Y., Ogaki, K., Hattori, N., & Aoki, S. (2020). MR Biomarkers of Degenerative Brain Disorders Derived From Diffusion Imaging. *Journal of Magnetic Resonance Imaging : JMRI*, *52*(6), 1620–1636. https://doi.org/10.1002/jmri.27019

Bartzokis, G. (2004). Age-related myelin breakdown: A developmental model of cognitive decline and Alzheimer's disease. *Neurobiology of Aging*, *25*(1), 5–18. https://doi.org/10.1016/j.neurobiolaging.2003.03.001

Beck, D., de Lange, A.-M. G., Maximov, I. I., Richard, G., Andreassen, O. A., Nordvik, J. E., & Westlye, L. T. (2021). White matter microstructure across the adult lifespan: A mixed longitudinal and cross-sectional study using advanced diffusion models and brain-age prediction. *NeuroImage*, *224*, 117441. https://doi.org/10.1016/j.neuroimage.2020.117441

Benitez, A., Fieremans, E., Jensen, J. H., Falangola, M. F., Tabesh, A., Ferris, S. H., & Helpern, J. A. (2014). White matter tract integrity metrics reflect the vulnerability of late-myelinating tracts in Alzheimer's disease. *NeuroImage. Clinical*, *4*, 64–71. https://doi.org/10.1016/j.nicl.2013.11.001

Billiet, T., Vandenbulcke, M., Mädler, B., Peeters, R., Dhollander, T., Zhang, H., Deprez, S., Van den Bergh, B. R. H., Sunaert, S., & Emsell, L. (2015). Age-related microstructural differences quantified using myelin water imaging and advanced diffusion MRI. *Neurobiology of Aging*, *36*(6), 2107–2121. https://doi.org/10.1016/j.neurobiolaging.2015.02.029

Cai, L. Y., Yang, Q., Hansen, C. B., Nath, V., Ramadass, K., Johnson, G. W., Conrad, B. N., Boyd, B. D., Begnoche, J. P., Beason-Held, L. L., Shafer, A. T., Resnick, S. M., Taylor, W. D., Price, G. R., Morgan, V. L., Rogers, B. P., Schilling, K. G., & Landman, B. A. (2021). PreQual: An automated pipeline for integrated preprocessing and quality assurance of diffusion weighted MRI images. *Magnetic Resonance in Medicine*, *86*(1), 456–470. https://doi.org/10.1002/mrm.28678

Chang, L.-C., Walker, L., & Pierpaoli, C. (2012). Informed RESTORE: A method for robust estimation of diffusion tensor from low redundancy datasets in the presence of physiological noise artifacts. *Magnetic Resonance in Medicine*, *68*(5), 1654–1663. https://doi.org/10.1002/mrm.24173

Cieslak, M., Cook, P. A., He, X., Yeh, F.-C., Dhollander, T., Adebimpe, A., Aguirre, G. K., Bassett, D. S., Betzel, R. F., Bourque, J., Cabral, L. M., Davatzikos, C., Detre, J. A., Earl, E., Elliott, M. A., Fadnavis, S., Fair, D. A., Foran, W., Fotiadis, P., … Satterthwaite, T. D. (2021). QSIPrep: An integrative platform for preprocessing and reconstructing diffusion MRI data. *Nature Methods*, *18*(7), 775–778. https://doi.org/10.1038/s41592-021-01185-5

Cieslak, M., Cook, P. A., Shafiei, G., Tapera, T. M., Radhakrishnan, H., Elliott, M., Roalf, D. R., Oathes, D. J., Bassett, D. S., Tisdall, M. D., Rokem, A., Grafton, S. T., & Satterthwaite, T. D. (2024). Diffusion MRI head motion correction methods are highly accurate but impacted by denoising and sampling scheme. *Human Brain Mapping*, *45*(2), e26570. https://doi.org/10.1002/hbm.26570

Collier, Q., Veraart, J., Jeurissen, B., den Dekker, A. J., & Sijbers, J. (2015). Iterative reweighted linear least squares for accurate, fast, and robust estimation of diffusion magnetic resonance parameters. *Magnetic Resonance in Medicine*, *73*(6), 2174–2184. https://doi.org/10.1002/mrm.25351

Cordero-Grande, L., Christiaens, D., Hutter, J., Price, A. N., & Hajnal, J. V. (2019). Complex diffusion-weighted image estimation via matrix recovery under general noise models. *NeuroImage*, *200*, 391–404. https://doi.org/10.1016/j.neuroimage.2019.06.039

Cui, Z., Zhong, S., Xu, P., Gong, G., & He, Y. (2013). PANDA: a pipeline toolbox for analyzing brain diffusion images. *Frontiers in Human Neuroscience*, *7*. https://doi.org/10.3389/fnhum.2013.00042

Dietrich, O., Heiland, S., & Sartor, K. (2001). Noise correction for the exact determination of apparent diffusion coefficients at low SNR. *Magnetic Resonance in Medicine*, *45*(3), 448–453. https://doi.org/10.1002/1522-2594(200103)45:3<448::AID-MRM1059>3.0.CO;2-W

Dubkov, A. A., & Malakhov, A. N. (1976). Properties and interdependence of the cumulants of a random variable. *Radiophysics and Quantum Electronics*, *19*(8), 833–839. https://doi.org/10.1007/BF01043479

Fadnavis, S., Batson, J., & Garyfallidis, E. (2020). Patch2Self: Denoising Diffusion MRI with Self-Supervised Learning. *CoRR*, *abs/2011.01355*. https://arxiv.org/abs/2011.01355

Garyfallidis, E., Brett, M., Amirbekian, B., Rokem, A., van der Walt, S., Descoteaux, M., & Nimmo-Smith, I. (2014). Dipy, a library for the analysis of diffusion MRI data. *Frontiers in Neuroinformatics*, *8*, 8. https://doi.org/10.3389/fninf.2014.00008

Gavish, M., & Donoho, D. L. (2017). Optimal Shrinkage of Singular Values. *IEEE Trans. Inf. Theor.*, *63*(4), 2137–2152. https://doi.org/10.1109/TIT.2017.2653801

Glasser, M. F., Sotiropoulos, S. N., Wilson, J. A., Coalson, T. S., Fischl, B., Andersson, J. L., Xu, J., Jbabdi, S., Webster, M., Polimeni, J. R., Van Essen, D. C., & Jenkinson, M. (2013). The minimal preprocessing pipelines for the Human Connectome Project. *NeuroImage*, *80*, 105–124. https://doi.org/10.1016/j.neuroimage.2013.04.127

Henriques, R. N., Correia, M. M., Marrale, M., Huber, E., Kruper, J., Koudoro, S., Yeatman, J. D., Garyfallidis, E., & Rokem, A. (2021). Diffusional Kurtosis Imaging in the Diffusion Imaging in Python Project. *Frontiers in Human Neuroscience*, *15*, 675433. https://doi.org/10.3389/fnhum.2021.675433

Henriques, R. N., Jespersen, S. N., Jones, D. K., & Veraart, J. (2021). Toward more robust and reproducible diffusion kurtosis imaging. *Magnetic Resonance in Medicine*, *86*(3), 1600–1613. https://doi.org/10.1002/mrm.28730

Hua, K., Zhang, J., Wakana, S., Jiang, H., Li, X., Reich, D. S., Calabresi, P. A., Pekar, J. J., van Zijl, P. C. M., & Mori, S. (2008). Tract probability maps in stereotaxic spaces: Analyses of white matter anatomy and tract-specific quantification. *NeuroImage*, *39*(1), 336–347. https://doi.org/10.1016/j.neuroimage.2007.07.053

Irfanoglu, M. O., Nayak, A., Jenkins, J., & Pierpaoli, C. (2017). *TORTOISE v3: Improvements and new features of the NIH diffusion MRI processing pipeline*. Program and proceedings of the ISMRM 25th annual meeting and exhibition, Honolulu, HI, USA.

Jain, A. K. (1989). *Fundamentals of digital image processing*. Prentice-Hall, Inc.

Jenkinson, M., Bannister, P., Brady, M., & Smith, S. (2002). Improved optimization for the robust and accurate linear registration and motion correction of brain images. *NeuroImage*, *17*(2), 825–841. https://doi.org/10.1016/s1053-8119(02)91132-8

Jenkinson, M., & Smith, S. (2001). A global optimisation method for robust affine registration of brain images. *Medical Image Analysis*, *5*(2), 143–156. https://doi.org/10.1016/s1361-8415(01)00036-6

Jensen, J. H., & Helpern, J. A. (2010). MRI quantification of non-Gaussian water diffusion by kurtosis analysis. *NMR in Biomedicine*, *23*(7), 698–710. https://doi.org/10.1002/nbm.1518

Jones, D. K., & Cercignani, M. (2010). Twenty-five pitfalls in the analysis of diffusion MRI data. *NMR in Biomedicine*, *23*(7), 803–820. https://doi.org/10.1002/nbm.1543

Kellner, E., Dhital, B., Kiselev, V. G., & Reisert, M. (2016). Gibbs-ringing artifact removal based on local subvoxel-shifts. *Magnetic Resonance in Medicine*, *76*(5), 1574–1581. https://doi.org/10.1002/mrm.26054

Koay, C. G., & Basser, P. J. (2006). Analytically exact correction scheme for signal extraction from noisy magnitude MR signals. *Journal of Magnetic Resonance (San Diego, Calif. : 1997)*, *179*(2), 317–322. https://doi.org/10.1016/j.jmr.2006.01.016

Kodiweera, C., Alexander, A. L., Harezlak, J., McAllister, T. W., & Wu, Y.-C. (2016). Age effects and sex differences in human brain white matter of young to middle-aged adults: A DTI, NODDI, and q-space study. *NeuroImage*, *128*, 180–192. https://doi.org/10.1016/j.neuroimage.2015.12.033

Kornaropoulos, E. N., Winzeck, S., Rumetshofer, T., Wikstrom, A., Knutsson, L., Correia, M. M., Sundgren, P. C., & Nilsson, M. (2022). Sensitivity of Diffusion MRI to White Matter Pathology: Influence of Diffusion Protocol, Magnetic Field Strength, and Processing Pipeline in Systemic Lupus Erythematosus. *Frontiers in Neurology*, *13*. https://doi.org/10.3389/fneur.2022.837385

Kuder, T. A., Stieltjes, B., Bachert, P., Semmler, W., & Laun, F. B. (2012). Advanced fit of the diffusion kurtosis tensor by directional weighting and regularization. *Magnetic Resonance in Medicine*, *67*(5), 1401–1411. https://doi.org/10.1002/mrm.23133

Le Bihan, D., Poupon, C., Amadon, A., & Lethimonnier, F. (2006). Artifacts and pitfalls in diffusion MRI. *Journal of Magnetic Resonance Imaging : JMRI*, *24*(3), 478–488. https://doi.org/10.1002/jmri.20683

Lee, H.-H., Novikov, D. S., & Fieremans, E. (2021). Removal of partial Fourier-induced Gibbs (RPG) ringing artifacts in MRI. *Magnetic Resonance in Medicine*, *86*(5), 2733–2750. https://doi.org/10.1002/mrm.28830

Maximov, I. I., Alnæs, D., & Westlye, L. T. (2019). Towards an optimised processing pipeline for diffusion MRI data: Effects of artefact corrections on diffusion metrics and their age associations in UK Biobank. *bioRxiv*. https://doi.org/10.1101/511964

Moura, L. M., Luccas, R., de Paiva, J. P. Q., Amaro, E. J., Leemans, A., Leite, C. da C., Otaduy, M. C. G., & Conforto, A. B. (2019). Diffusion Tensor Imaging Biomarkers to Predict Motor Outcomes in Stroke: A Narrative Review. *Frontiers in Neurology*, *10*, 445. https://doi.org/10.3389/fneur.2019.00445

Novikov, D. S., Fieremans, E., Jespersen, S. N., & Kiselev, V. G. (2019). Quantifying brain microstructure with diffusion MRI: Theory and parameter estimation. *NMR in Biomedicine*, *32*(4), e3998. https://doi.org/10.1002/nbm.3998

Olson, D. V., Arpinar, V. E., & Muftuler, L. T. (2018). Assessing diffusion kurtosis tensor estimation methods using a digital brain phantom derived from human connectome project data. *Magnetic Resonance Imaging*, *48*, 122–128. https://doi.org/10.1016/j.mri.2017.12.026

Ouyang, Y., Cui, D., Yuan, Z., Liu, Z., Jiao, Q., Yin, T., & Qiu, J. (2021). Analysis of Age-Related White Matter Microstructures Based on Diffusion Tensor Imaging. *Frontiers in Aging Neuroscience*, *13*. https://doi.org/10.3389/fnagi.2021.664911

Pang, M., Chen, J., Ades-Aron, B., Shepherd, T., Osorio, R., & Fieremans, E. (2022). Identifying Sex Differences in Aging with Diffusion MRI. *American Geriatrics Society*, B191.

Paydar, A., Fieremans, E., Nwankwo, J. I., Lazar, M., Sheth, H. D., Adisetiyo, V., Helpern, J. A., Jensen, J. H., & Milla, S. S. (2014). Diffusional Kurtosis Imaging of the Developing Brain. *American Journal of Neuroradiology*, *35*(4), 808–814. https://doi.org/10.3174/ajnr.A3764

Perrone, D., Aelterman, J., Pižurica, A., Jeurissen, B., Philips, W., & Leemans, A. (2015). The effect of Gibbs ringing artifacts on measures derived from diffusion MRI. *NeuroImage*, *120*, 441–455. https://doi.org/10.1016/j.neuroimage.2015.06.068

Schilling, K. G., Archer, D., Yeh, F.-C., Rheault, F., Cai, L. Y., Hansen, C., Yang, Q., Ramdass, K., Shafer, A. T., Resnick, S. M., Pechman, K. R., Gifford, K. A., Hohman, T. J., Jefferson, A., Anderson, A. W., Kang, H., & Landman, B. A. (2022). Aging and white matter microstructure and macrostructure: A longitudinal multi-site diffusion MRI study of 1218 participants. *Brain Structure and Function*. https://doi.org/10.1007/s00429-022-02503-z

Shepp, L. A., & Logan, B. F. (1974). The Fourier reconstruction of a head section. *IEEE Transactions on Nuclear Science*, *21*(3), 21–43. https://doi.org/10.1109/TNS.1974.6499235

Smith, S. M., Jenkinson, M., Woolrich, M. W., Beckmann, C. F., Behrens, T. E. J., Johansen-Berg, H., Bannister, P. R., Luca, M. D., Drobnjak, I., Flitney, D., Niazy, R. K., Saunders, J., Vickers, J., Zhang, Y., Stefano, N. D., Brady, J., & Matthews, P. M. (2004). Advances in functional and structural MR image analysis and implementation as FSL. *NeuroImage*, *23*, S208–S219.

Taha, H. T., Chad, J. A., & Chen, J. J. (2022). DKI enhances the sensitivity and interpretability of age-related DTI patterns in the white matter of UK biobank participants. *Neurobiology of Aging*, *115*, 39–49. https://doi.org/10.1016/j.neurobiolaging.2022.03.008

Toschi, N., Gisbert, R. A., Passamonti, L., Canals, S., & Santis, S. D. (2020). Multishell diffusion imaging reveals sex-specific trajectories of early white matter degeneration in normal aging. *Neurobiology of Aging*, *86*, 191–200. https://doi.org/10.1016/j.neurobiolaging.2019.11.014

Tournier, J.-D., Smith, R., Raffelt, D., Tabbara, R., Dhollander, T., Pietsch, M., Christiaens, D., Jeurissen, B., Yeh, C.-H., & Connelly, A. (2019). MRtrix3: A fast, flexible and open software framework for medical image processing and visualisation. *NeuroImage*, *202*, 116137. https://doi.org/10.1016/j.neuroimage.2019.116137

Tristán-Vega, A., Aja-Fernández, S., & Westin, C.-F. (2012). Least squares for diffusion tensor estimation revisited: Propagation of uncertainty with Rician and non-Rician signals. *NeuroImage*, *59*(4), 4032–4043. https://doi.org/10.1016/j.neuroimage.2011.09.074

Veraart, J., Fieremans, E., Jelescu, I. O., Knoll, F., & Novikov, D. S. (2016). Gibbs ringing in diffusion MRI. *Magnetic Resonance in Medicine*, *76*(1), 301–314. https://doi.org/10.1002/mrm.25866

Veraart, J., Novikov, D. S., Christiaens, D., Ades-Aron, B., Sijbers, J., & Fieremans, E. (2016). Denoising of diffusion MRI using random matrix theory. *NeuroImage*, *142*, 394–406. https://doi.org/10.1016/j.neuroimage.2016.08.016

Veraart, J., Sijbers, J., Sunaert, S., Leemans, A., & Jeurissen, B. (2013). Weighted linear least squares estimation of diffusion MRI parameters: Strengths, limitations, and pitfalls. *NeuroImage*, *81*, 335–346. https://doi.org/10.1016/j.neuroimage.2013.05.028

Vos, S. B., Tax, C. M. W., Luijten, P. R., Ourselin, S., Leemans, A., & Froeling, M. (2017). The importance of correcting for signal drift in diffusion MRI. *Magnetic Resonance in Medicine*, *77*(1), 285–299. https://doi.org/10.1002/mrm.26124

Yeatman, J. D., Wandell, B. A., & Mezer, A. A. (2014). Lifespan maturation and degeneration of human brain white matter. *Nature Communications*, *5*(1), 4932. https://doi.org/10.1038/ncomms5932

**Table S1.** Adjusted $R^2$ from quadratic and linear fit of age correlation with DTI and DKI parameters in white matter regions (median value) using no preprocessing pipeline, E+M, DESIGNER-Dv1, and DESIGNER-Dv2 preprocessing pipeline (Prisma, TE = 70ms, N = 142). Adjusted $R^2$s from quadratic fit are larger than adjusted $R^2$s from linear fit for all age associations. Bolded adjusted $R^2$ indicates statistically significant age association with adjusted $R^2 > 0.1$. Cell color-scale from light to dark blue indicates lowest to highest adjusted $R^2$.

| | | | MD | RD | AD | FA | MK | RK | AK |
|---|---|---|---|---|---|---|---|---|---|
| PLIC | None | Quadratic | 0.00 | 0.07 | **0.15** | **0.12** | 0.01 | 0.01 | 0.02 |
| | | Linear | 0.00 | 0.06 | **0.14** | **0.11** | -0.01 | -0.01 | 0.02 |
| | E+M | Quadratic | 0.06 | 0.05 | **0.16** | 0.10 | 0.02 | 0.04 | 0.08 |
| | | Linear | 0.03 | 0.05 | **0.15** | 0.10 | 0.02 | 0.04 | 0.08 |
| | DESIGNER-Dv1 | Quadratic | 0.03 | 0.04 | **0.12** | 0.08 | 0.03 | 0.08 | 0.08 |
| | | Linear | 0.02 | 0.03 | **0.11** | 0.08 | 0.03 | 0.08 | 0.07 |
| | DESIGNER-Dv2 | Quadratic | 0.03 | 0.05 | 0.10 | 0.09 | 0.02 | 0.08 | 0.10 |
| | | Linear | 0.01 | 0.04 | 0.09 | 0.09 | 0.02 | 0.08 | 0.09 |
| SCC | None | Quadratic | 0.01 | 0.05 | 0.02 | 0.06 | 0.01 | 0.02 | 0.08 |
| | | Linear | 0.01 | 0.05 | 0.01 | 0.06 | 0.01 | 0.02 | 0.05 |
| | E+M | Quadratic | 0.00 | 0.03 | 0.01 | 0.05 | 0.00 | 0.00 | **0.13** |
| | | Linear | 0.00 | 0.03 | 0.01 | 0.05 | 0.00 | 0.00 | **0.11** |
| | DESIGNER-Dv1 | Quadratic | 0.01 | 0.04 | 0.00 | 0.05 | 0.00 | -0.01 | **0.16** |
| | | Linear | 0.01 | 0.04 | 0.00 | 0.05 | -0.01 | -0.01 | **0.11** |
| | DESIGNER-Dv2 | Quadratic | 0.04 | 0.09 | 0.00 | 0.08 | 0.00 | 0.00 | **0.16** |
| | | Linear | 0.04 | 0.08 | 0.00 | 0.08 | -0.01 | 0.00 | **0.11** |
| GCC | None | Quadratic | **0.17** | **0.21** | 0.07 | **0.20** | 0.01 | 0.01 | 0.02 |
| | | Linear | **0.13** | **0.19** | 0.01 | **0.19** | -0.01 | -0.01 | 0.02 |
| | E+M | Quadratic | **0.19** | **0.22** | 0.05 | **0.18** | 0.03 | 0.01 | 0.00 |
| | | Linear | **0.14** | **0.20** | -0.01 | **0.18** | 0.01 | 0.00 | 0.00 |
| | DESIGNER-Dv1 | Quadratic | **0.19** | **0.24** | 0.02 | **0.22** | 0.07 | 0.06 | 0.01 |
| | | Linear | **0.16** | **0.22** | -0.01 | **0.21** | 0.04 | 0.03 | -0.01 |
| | DESIGNER-Dv2 | Quadratic | **0.22** | **0.29** | 0.01 | **0.24** | **0.11** | 0.08 | 0.00 |
| | | Linear | **0.18** | **0.26** | 0.00 | **0.23** | 0.07 | 0.06 | -0.01 |
| ACR | None | Quadratic | **0.24** | **0.34** | 0.00 | **0.27** | 0.05 | 0.03 | 0.01 |
| | | Linear | **0.21** | **0.31** | 0.00 | **0.26** | 0.03 | 0.02 | 0.00 |
| | E+M | Quadratic | **0.23** | **0.32** | 0.00 | **0.29** | **0.12** | **0.13** | 0.00 |
| | | Linear | **0.20** | **0.30** | -0.01 | **0.28** | **0.11** | **0.12** | -0.01 |
| | DESIGNER-Dv1 | Quadratic | **0.23** | **0.32** | -0.01 | **0.29** | **0.14** | **0.19** | 0.01 |
| | | Linear | **0.21** | **0.29** | -0.01 | **0.26** | **0.13** | **0.17** | 0.00 |
| | DESIGNER-Dv2 | Quadratic | **0.26** | **0.33** | 0.00 | **0.29** | **0.16** | **0.23** | 0.01 |
| | | Linear | **0.23** | **0.30** | -0.01 | **0.25** | **0.14** | **0.21** | 0.00 |

**Table S2.** Bonferroni adjusted P-values for age correlation with DTI and DKI parameters in white matter regions (median value) using no preprocessing pipeline, E+M, DESIGNER-Dv1, and DESIGNER-Dv2 preprocessing pipeline (Prisma, TE = 70ms, N = 142).

|  |  | MD | RD | AD | FA | MK | RK | AK |
|---|---|---|---|---|---|---|---|---|
| PLIC | None | 1.000 | 0.037 | <0.001 | <0.001 | 1.000 | 1.000 | 1.000 |
|  | E+M | 0.051 | 0.104 | <0.001 | 0.002 | 1.000 | 0.318 | 0.014 |
|  | DESIGNER-Dv1 | 0.615 | 0.416 | <0.001 | 0.007 | 0.489 | 0.012 | 0.016 |
|  | DESIGNER-Dv2 | 0.681 | 0.174 | 0.003 | 0.005 | 1.000 | 0.010 | 0.003 |
| SCC | None | 1.000 | 0.128 | 1.000 | 0.061 | 1.000 | 1.000 | 0.014 |
|  | E+M | 1.000 | 0.487 | 1.000 | 0.130 | 1.000 | 1.000 | <0.001 |
|  | DESIGNER-Dv1 | 1.000 | 0.297 | 1.000 | 0.119 | 1.000 | 1.000 | <0.001 |
|  | DESIGNER-Dv2 | 0.223 | 0.006 | 1.000 | 0.008 | 1.000 | 1.000 | <0.001 |
| GCC | None | <0.001 | <0.001 | 0.019 | <0.001 | 1.000 | 1.000 | 1.000 |
|  | E+M | <0.001 | <0.001 | 0.173 | <0.001 | 0.543 | 1.000 | 1.000 |
|  | DESIGNER-Dv1 | <0.001 | <0.001 | 1.000 | <0.001 | 0.025 | 0.066 | 1.000 |
|  | DESIGNER-Dv2 | <0.001 | <0.001 | 1.000 | <0.001 | 0.001 | 0.008 | 1.000 |
| ACR | None | <0.001 | <0.001 | 1.000 | <0.001 | 0.174 | 0.513 | 1.000 |
|  | E+M | <0.001 | <0.001 | 1.000 | <0.001 | <0.001 | <0.001 | 1.000 |
|  | DESIGNER-Dv1 | <0.001 | <0.001 | 1.000 | <0.001 | <0.001 | <0.001 | 1.000 |
|  | DESIGNER-Dv2 | <0.001 | <0.001 | 1.000 | <0.001 | <0.001 | <0.001 | 1.000 |

**Table S3.** Bonferroni adjusted P-values for age correlation with DTI and DKI parameters in white matter regions (median value) using no preprocessing pipeline, E+M, DESIGNER-Dv1, and DESIGNER-Dv2 preprocessing pipeline (Prisma, TE = 95ms, N = 120).

|  |  | MD | RD | AD | FA | MK | RK | AK |
|---|---|---|---|---|---|---|---|---|
| PLIC | None | 1.000 | 1.000 | 0.388 | 1.000 | 0.855 | 1.000 | 0.282 |
|  | E+M | 1.000 | 1.000 | 1.000 | 1.000 | 0.948 | 1.000 | 0.169 |
|  | DESIGNER-Dv1 | 1.000 | 1.000 | 1.000 | 1.000 | 0.064 | 0.134 | 0.110 |
|  | DESIGNER-Dv2 | 1.000 | 1.000 | 1.000 | 1.000 | 0.009 | 0.055 | 0.076 |
| SCC | None | 1.000 | 1.000 | 0.662 | 1.000 | 1.000 | 1.000 | 0.252 |
|  | E+M | 1.000 | 1.000 | 1.000 | 1.000 | 1.000 | 1.000 | 0.176 |
|  | DESIGNER-Dv1 | 1.000 | 1.000 | 1.000 | 1.000 | 1.000 | 1.000 | 0.023 |
|  | DESIGNER-Dv2 | 1.000 | 0.107 | 1.000 | 0.037 | 0.895 | 1.000 | 0.036 |
| GCC | None | 0.004 | <0.001 | 0.943 | <0.001 | 1.000 | 1.000 | 1.000 |
|  | E+M | 0.052 | 0.001 | 0.012 | <0.001 | 0.209 | 1.000 | 1.000 |
|  | DESIGNER-Dv1 | 0.059 | <0.001 | 0.139 | <0.001 | <0.001 | 0.001 | 1.000 |
|  | DESIGNER-Dv2 | 0.001 | <0.001 | 0.126 | <0.001 | <0.001 | <0.001 | 1.000 |
| ACR | None | <0.001 | <0.001 | 1.000 | <0.001 | 0.020 | 0.147 | 1.000 |
|  | E+M | <0.001 | <0.001 | 1.000 | <0.001 | 0.001 | 0.009 | 1.000 |
|  | DESIGNER-Dv1 | <0.001 | <0.001 | 1.000 | <0.001 | <0.001 | <0.001 | 0.513 |
|  | DESIGNER-Dv2 | <0.001 | <0.001 | 1.000 | <0.001 | <0.001 | <0.001 | 0.441 |

**Table S4.** Adjusted $R^2$ for age correlation with DTI and DKI parameters in white matter regions (median value) using no preprocessing pipeline, E+M, DESIGNER-Dv1, and DESIGNER-Dv2 preprocessing pipeline (Prisma, TE = 95ms, N = 120). Bolded adjusted $R^2$ indicates statistical significance with adjusted $R^2 > 0.1$. Cell color-scale from light to dark blue indicates lowest to highest adjusted $R^2$.

| | | MD | RD | AD | FA | MK | RK | AK |
|---|---|---|---|---|---|---|---|---|
| PLIC | None | 0.01 | -0.01 | 0.04 | 0.00 | 0.03 | 0.03 | 0.05 |
| | E+M | 0.00 | -0.01 | 0.01 | 0.00 | 0.03 | 0.02 | 0.05 |
| | DESIGNER-Dv1 | 0.01 | -0.01 | 0.01 | 0.00 | 0.07 | 0.06 | 0.06 |
| | DESIGNER-Dv2 | 0.00 | 0.00 | 0.01 | 0.02 | 0.10 | 0.07 | 0.07 |
| SCC | None | 0.01 | 0.02 | 0.03 | 0.02 | 0.02 | 0.01 | 0.05 |
| | E+M | -0.01 | 0.01 | 0.01 | 0.02 | 0.02 | 0.01 | 0.05 |
| | DESIGNER-Dv1 | -0.01 | 0.01 | 0.02 | 0.03 | 0.02 | 0.02 | 0.08 |
| | DESIGNER-Dv2 | 0.00 | 0.06 | 0.02 | 0.08 | 0.03 | 0.03 | 0.08 |
| GCC | None | **0.11** | **0.21** | 0.03 | **0.25** | 0.00 | 0.00 | 0.02 |
| | E+M | 0.07 | **0.13** | 0.09 | **0.20** | 0.05 | 0.01 | 0.00 |
| | DESIGNER-Dv1 | 0.07 | **0.19** | 0.06 | **0.26** | **0.14** | **0.13** | 0.02 |
| | DESIGNER-Dv2 | **0.12** | **0.25** | 0.06 | **0.31** | **0.17** | **0.18** | 0.03 |
| ACR | None | **0.24** | **0.29** | 0.02 | **0.26** | 0.09 | 0.06 | 0.00 |
| | E+M | **0.22** | **0.30** | 0.01 | **0.28** | **0.13** | **0.10** | 0.03 |
| | DESIGNER-Dv1 | **0.22** | **0.31** | 0.00 | **0.30** | **0.23** | **0.29** | 0.04 |
| | DESIGNER-Dv2 | **0.21** | **0.32** | 0.00 | **0.30** | **0.23** | **0.29** | 0.04 |

**Table S5.** Bonferroni adjusted P-values for age correlation with DTI and DKI parameters in white matter regions (median value) using no preprocessing pipeline, E+M, DESIGNER-Dv1, and DESIGNER-Dv2 preprocessing pipeline (Skyra, TE = 95ms, N = 262).

|  |  | MD | RD | AD | FA | MK | RK | AK |
|---|---|---|---|---|---|---|---|---|
| PLIC | None | 1.000 | 0.201 | 1.000 | 0.526 | 1.000 | 1.000 | <0.001 |
|  | E+M | 0.920 | 0.006 | 1.000 | 0.001 | 1.000 | 1.000 | <0.001 |
|  | DESIGNER-Dv1 | 1.000 | 0.041 | 1.000 | 0.007 | 1.000 | 1.000 | <0.001 |
|  | DESIGNER-Dv2 | 1.000 | 0.026 | 1.000 | 0.002 | 0.427 | 1.000 | <0.001 |
| SCC | None | 1.000 | 1.000 | 1.000 | 1.000 | 1.000 | 1.000 | 0.005 |
|  | E+M | 1.000 | 0.805 | 1.000 | 0.707 | 0.015 | 0.089 | <0.001 |
|  | DESIGNER-Dv1 | 1.000 | 1.316 | 1.000 | 1.000 | 0.114 | 0.084 | <0.001 |
|  | DESIGNER-Dv2 | 0.500 | 0.067 | 1.000 | 0.067 | 0.371 | 0.350 | <0.001 |
| GCC | None | <0.001 | <0.001 | 0.227 | <0.001 | 1.000 | 1.000 | 1.000 |
|  | E+M | <0.001 | <0.001 | 0.361 | <0.001 | 1.000 | 1.000 | 0.187 |
|  | DESIGNER-Dv1 | <0.001 | <0.001 | 0.549 | <0.001 | 0.001 | <0.001 | 0.104 |
|  | DESIGNER-Dv2 | <0.001 | <0.001 | 0.451 | <0.001 | <0.001 | <0.001 | 0.196 |
| ACR | None | <0.001 | <0.001 | 1.000 | <0.001 | 0.370 | 0.931 | 0.822 |
|  | E+M | <0.001 | <0.001 | 1.000 | <0.001 | 0.016 | 0.004 | 0.226 |
|  | DESIGNER-Dv1 | <0.001 | <0.001 | 1.000 | <0.001 | <0.001 | <0.001 | 0.613 |
|  | DESIGNER-Dv2 | <0.001 | <0.001 | 1.000 | <0.001 | <0.001 | <0.001 | 0.213 |

**Table S6.** Adjusted $R^2$ for age correlation with DTI and DKI parameters in white matter regions (median value) using no preprocessing pipeline, E+M, DESIGNER-Dv1, and DESIGNER-Dv2 preprocessing pipeline (Skyra, TE = 95ms, N = 262). Bolded adjusted $R^2$ indicates statistical significance with adjusted $R^2 > 0.1$. Cell color-scale from light to dark blue indicates lowest to highest adjusted $R^2$.

| | | MD | RD | AD | FA | MK | RK | AK |
|---|---|---|---|---|---|---|---|---|
| PLIC | None | 0.01 | 0.02 | 0.01 | 0.02 | 0.00 | 0.00 | 0.09 |
| | E+M | 0.01 | 0.05 | 0.00 | 0.06 | 0.00 | 0.00 | **0.12** |
| | DESIGNER-Dv1 | 0.01 | 0.03 | 0.00 | 0.05 | 0.01 | 0.00 | 0.07 |
| | DESIGNER-Dv2 | 0.01 | 0.04 | 0.00 | 0.05 | 0.02 | 0.00 | 0.08 |
| SCC | None | 0.00 | 0.00 | 0.01 | 0.01 | 0.00 | 0.00 | 0.05 |
| | E+M | 0.00 | 0.01 | 0.01 | 0.02 | 0.04 | 0.03 | 0.09 |
| | DESIGNER-Dv1 | 0.01 | 0.01 | 0.00 | 0.01 | 0.03 | 0.03 | **0.12** |
| | DESIGNER-Dv2 | 0.02 | 0.03 | 0.00 | 0.03 | 0.02 | 0.02 | 0.10 |
| GCC | None | **0.11** | **0.18** | 0.02 | **0.19** | 0.00 | 0.00 | 0.01 |
| | E+M | **0.10** | **0.20** | 0.02 | **0.20** | 0.00 | 0.00 | 0.02 |
| | DESIGNER-Dv1 | **0.10** | **0.20** | 0.02 | **0.22** | 0.06 | 0.07 | 0.03 |
| | DESIGNER-Dv2 | **0.15** | **0.24** | 0.02 | **0.25** | 0.08 | 0.08 | 0.02 |
| ACR | None | **0.17** | **0.26** | 0.01 | **0.22** | 0.02 | 0.01 | 0.01 |
| | E+M | **0.17** | **0.25** | 0.00 | **0.24** | 0.04 | 0.05 | 0.02 |
| | DESIGNER-Dv1 | **0.16** | **0.25** | 0.01 | **0.23** | 0.09 | **0.16** | 0.02 |
| | DESIGNER-Dv2 | **0.17** | **0.26** | 0.00 | **0.24** | **0.10** | **0.17** | 0.02 |

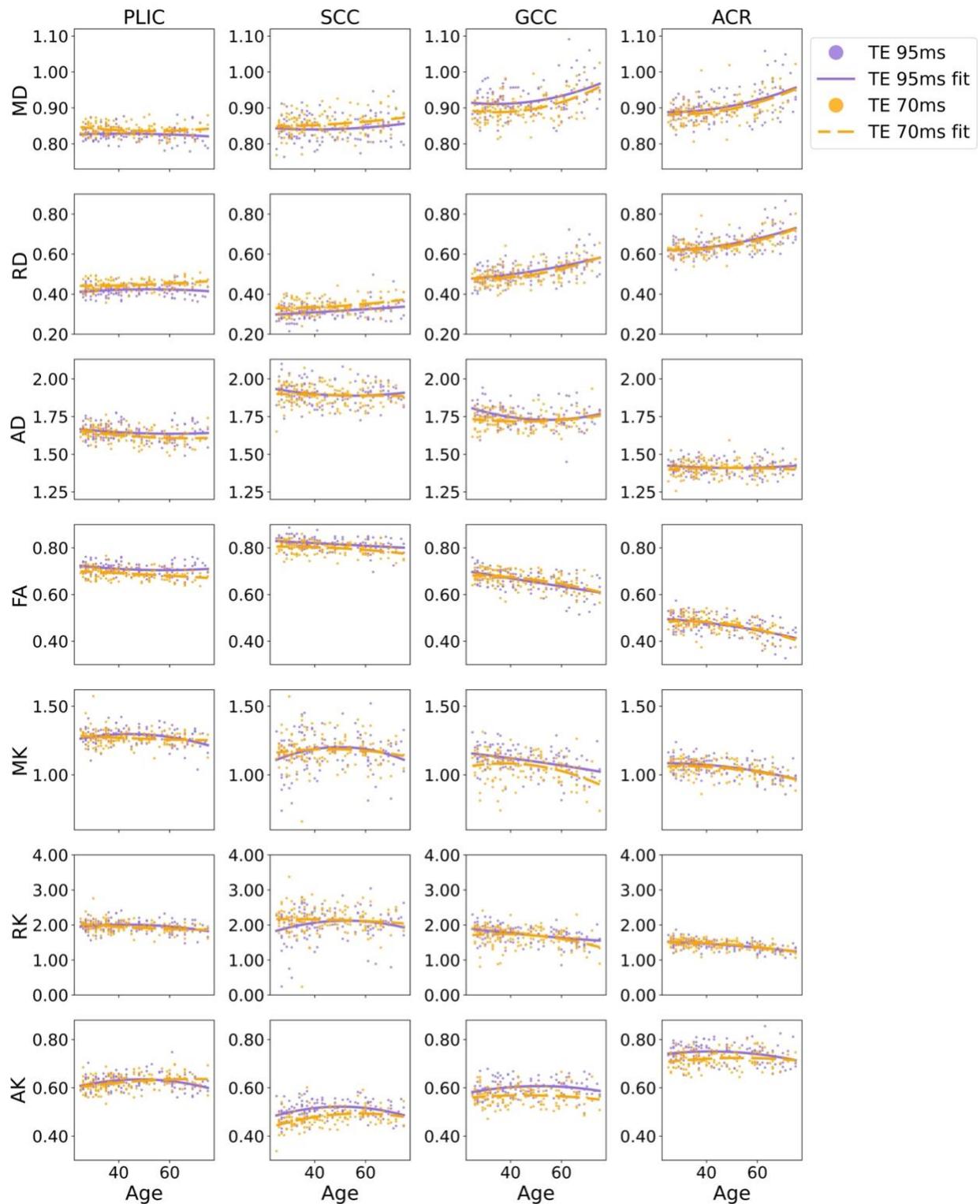

**Figure S1.** Age correlation with quadratic fits for DTI and DKI parameters in white matter ROIs (median) from DESIGNER-Dv2 pipeline for normal subjects with TE=95ms (N=120, Prisma) and TE=70ms (N=142, Prisma).

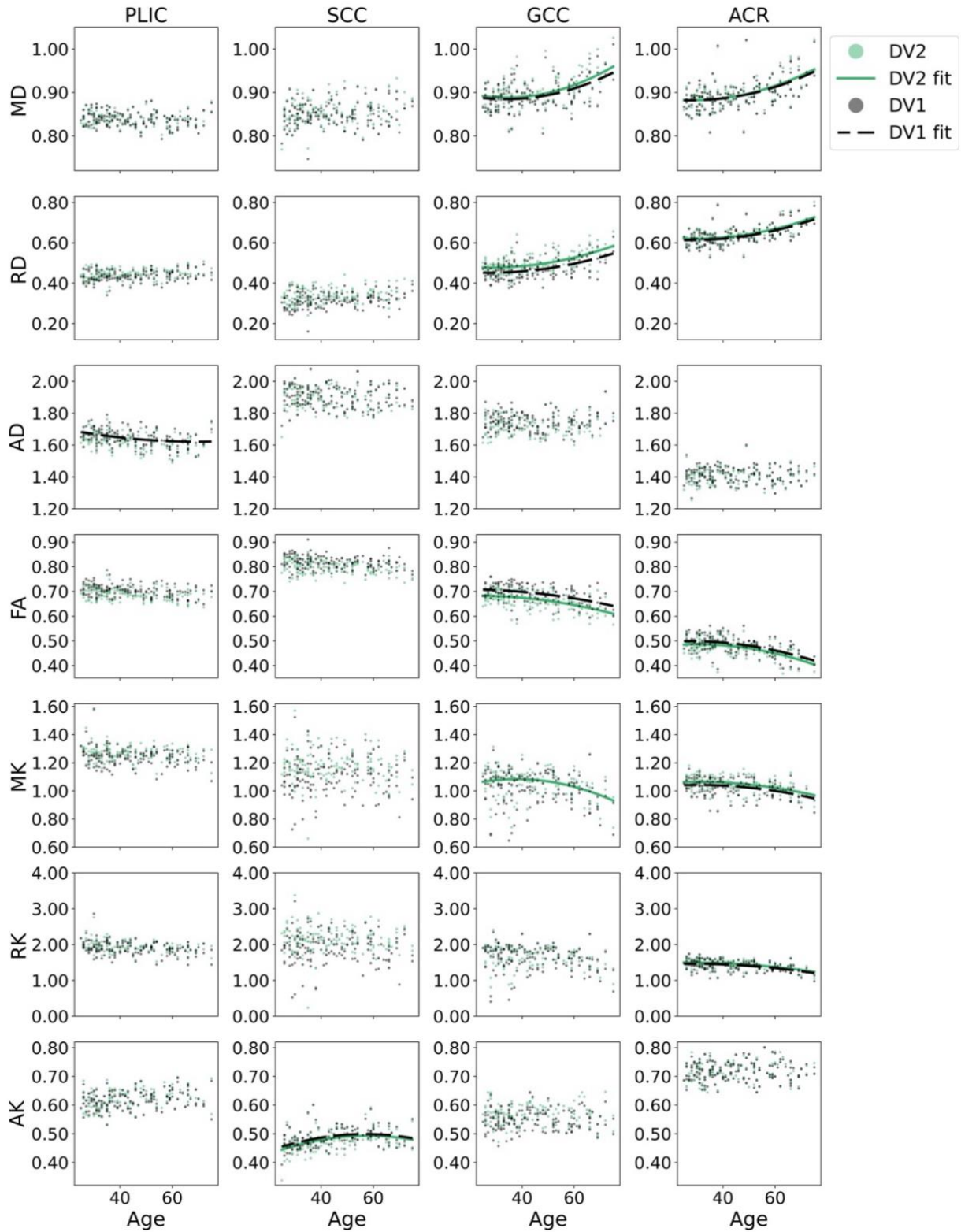

**Figure S2.** Age correlation with DTI and DKI parameters in white matter ROIs (median) from DESIGNER-Dv2 and DESIGNER-Dv1 pipeline of 142 healthy subjects (Prisma, TE = 70ms). Quadratic fits were plotted for statistically significant correlations with adjusted $R^2 > 0.1$ only.

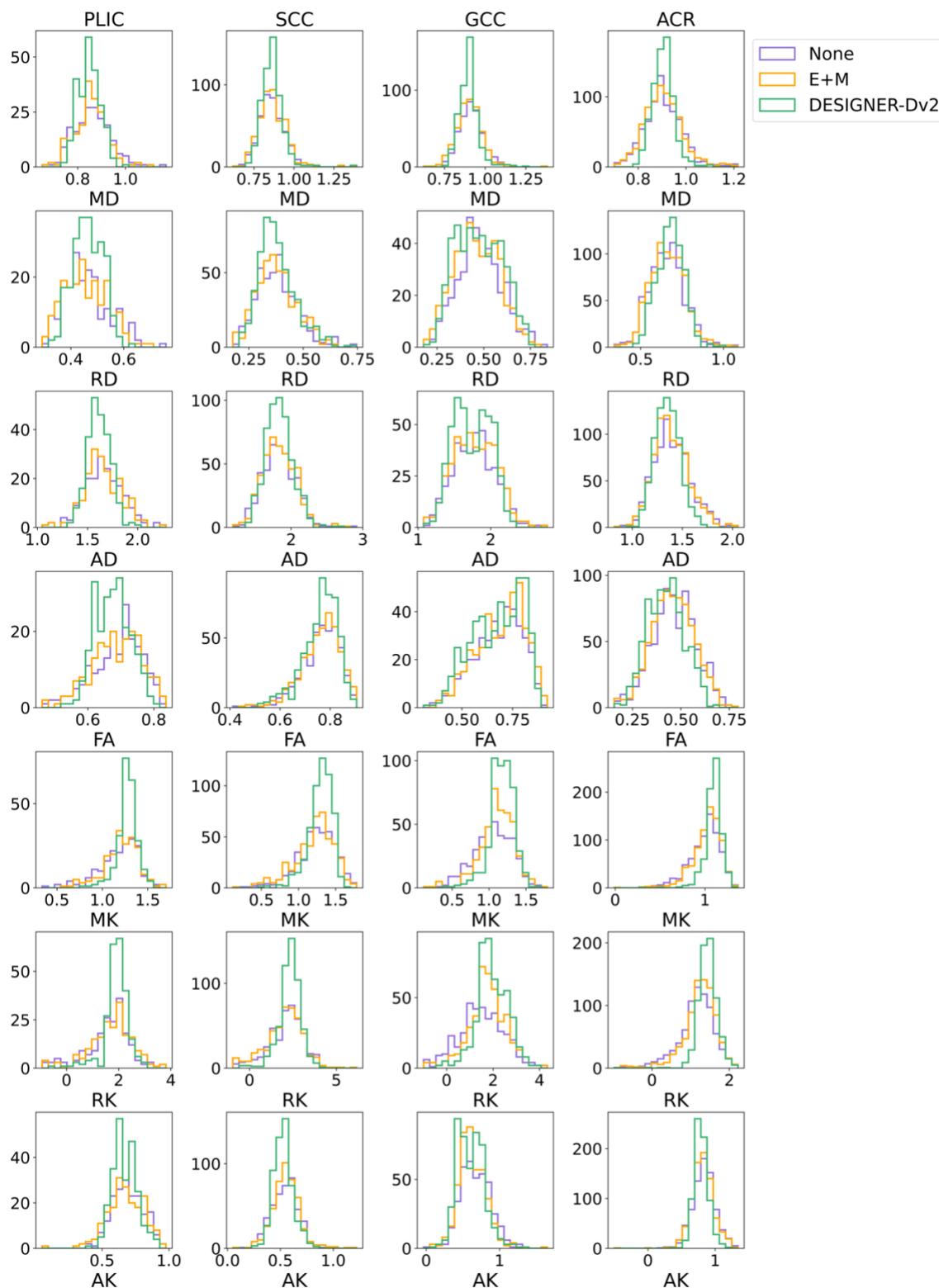

**Figure S3.** Probability density of DTI and DKI parameters after omitting outliers in white matter ROIs from data preprocessed with DESIGNER-Dv2, E+M, or no preprocessing pipeline for a healthy 60-year-old female.

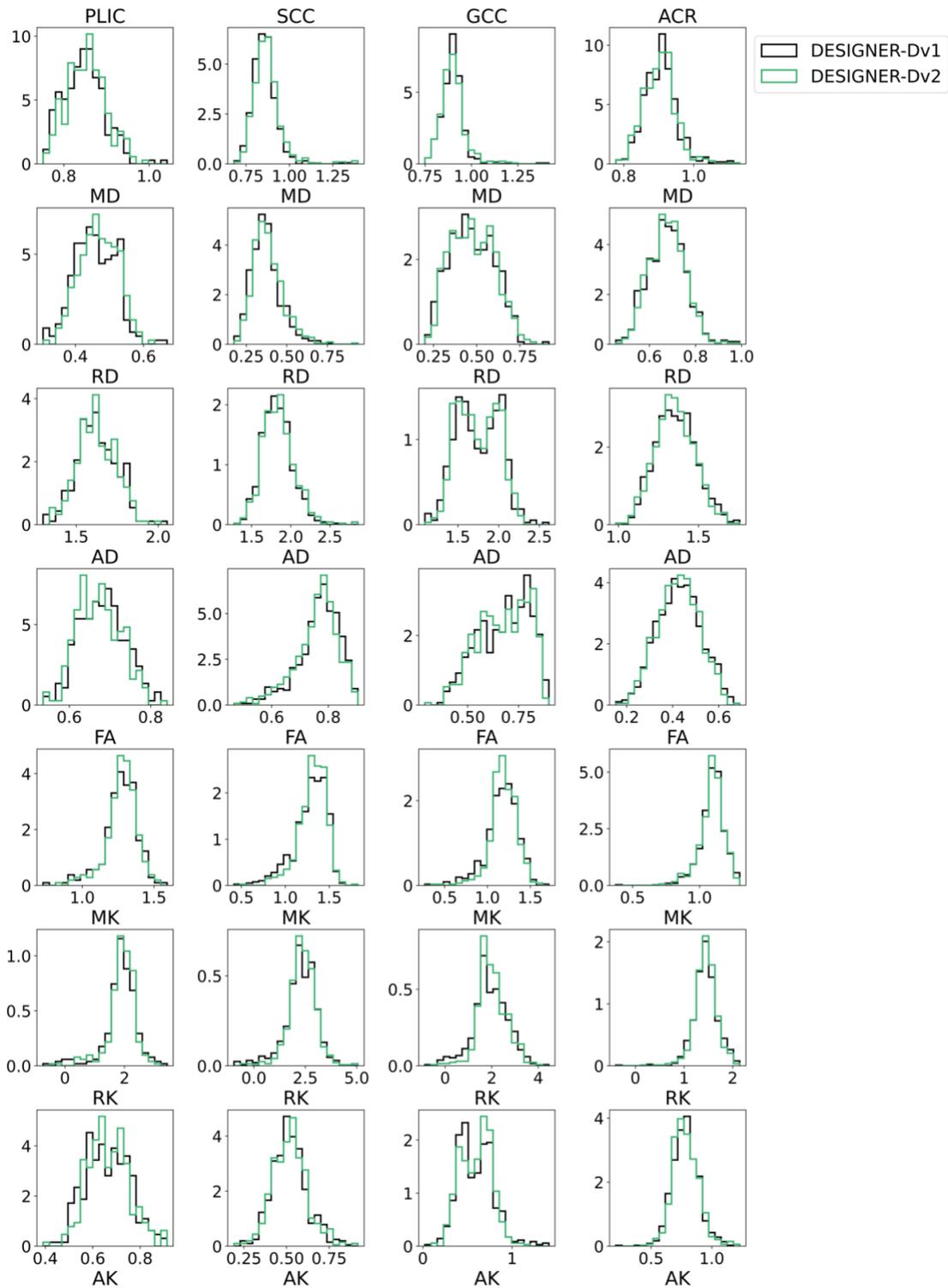

**Figure S4.** Probability density of DTI and DKI parameters after omitting outliers in white matter ROIs from data preprocessed with DESIGNER-Dv2 or DESIGNER-Dv1 pipeline, for a healthy 60-year-old female.

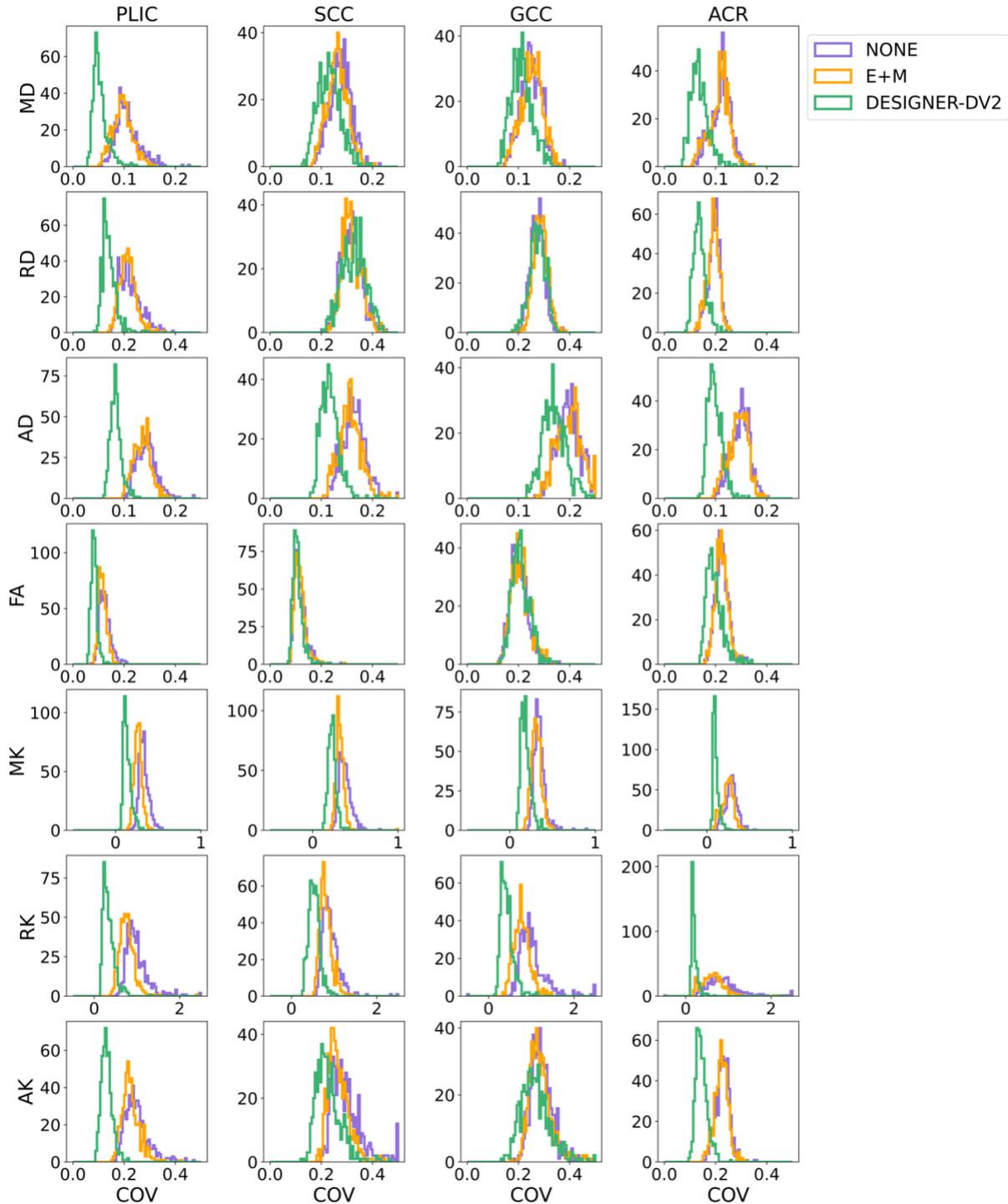

**Figure S5.** Histogram of coefficient of variation (COV=standard deviation/mean of ROI) after omitting outliers in white matter ROIs of DTI and DKI maps from DESIGNER-Dv2, E+M, or no preprocessing pipeline for 524 healthy subjects.

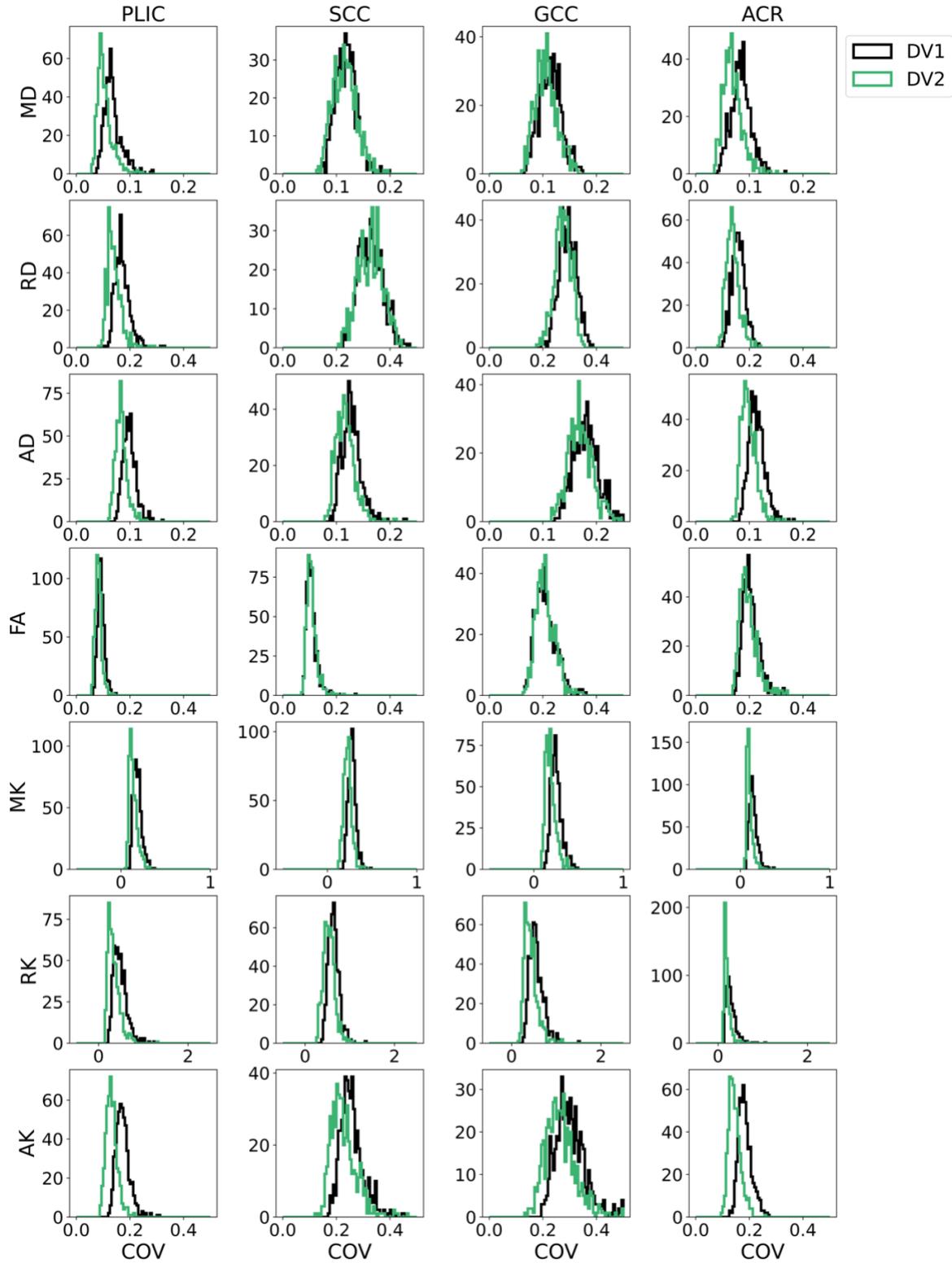

**Figure S6.** Histogram of coefficient of variation (COV=standard deviation/mean of ROI) after omitting outliers in white matter ROIs of DTI and DKI maps from DESIGNER-Dv2 and DESIGNER-Dv1 pipeline for 524 healthy subjects.

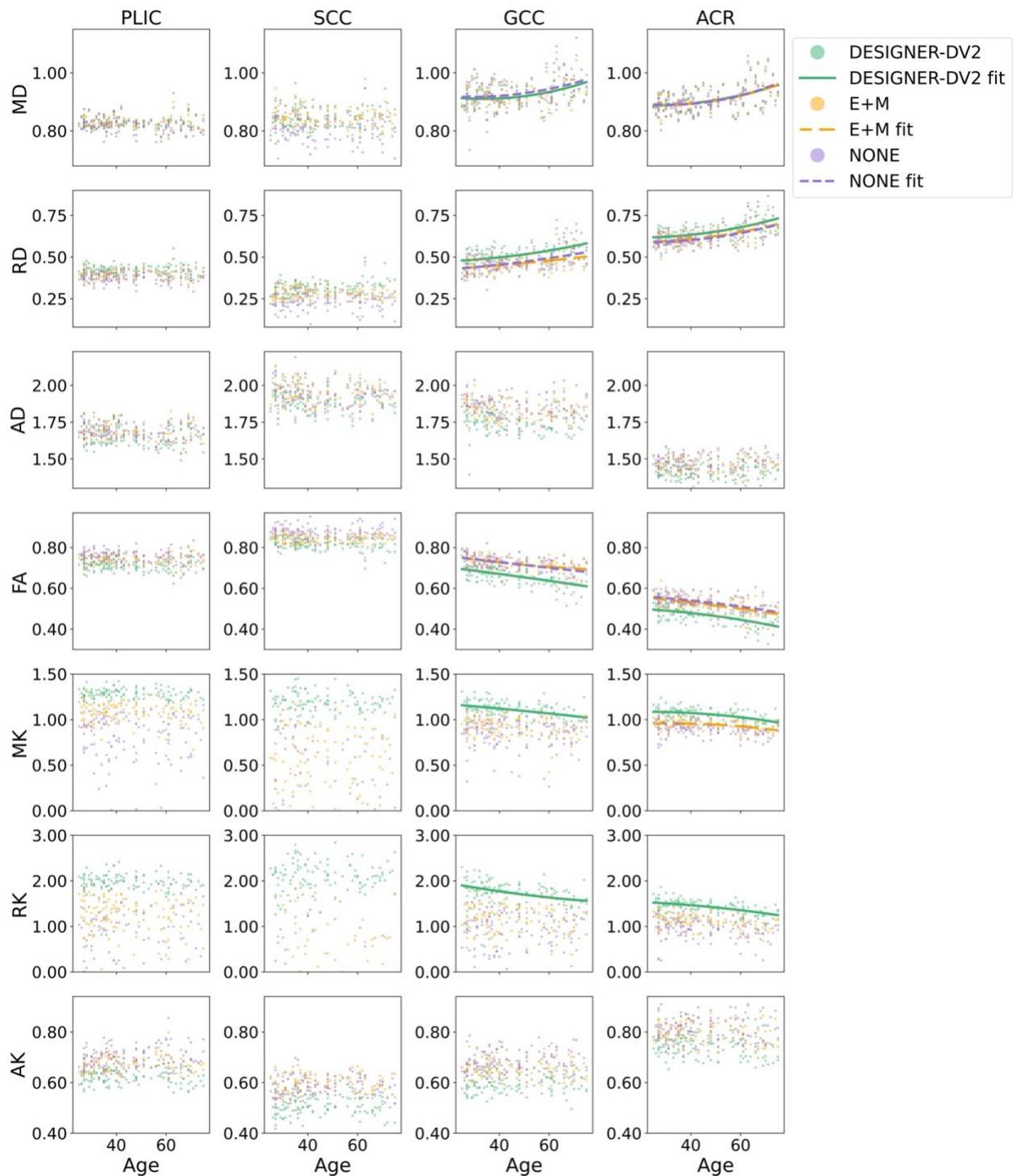

**Figure S7.** Age correlation with DTI and DKI parameters in white matter ROIs (median) from DESIGNER-Dv2 pipeline, E+M, or not preprocessing pipeline of 120 healthy subjects (Prisma, TE = 95ms). Quadratic fits were plotted for statistically significant correlations with adjusted $R^2 > 0.1$ only. MK and RK plots are zoomed in so not all datapoints are visible.

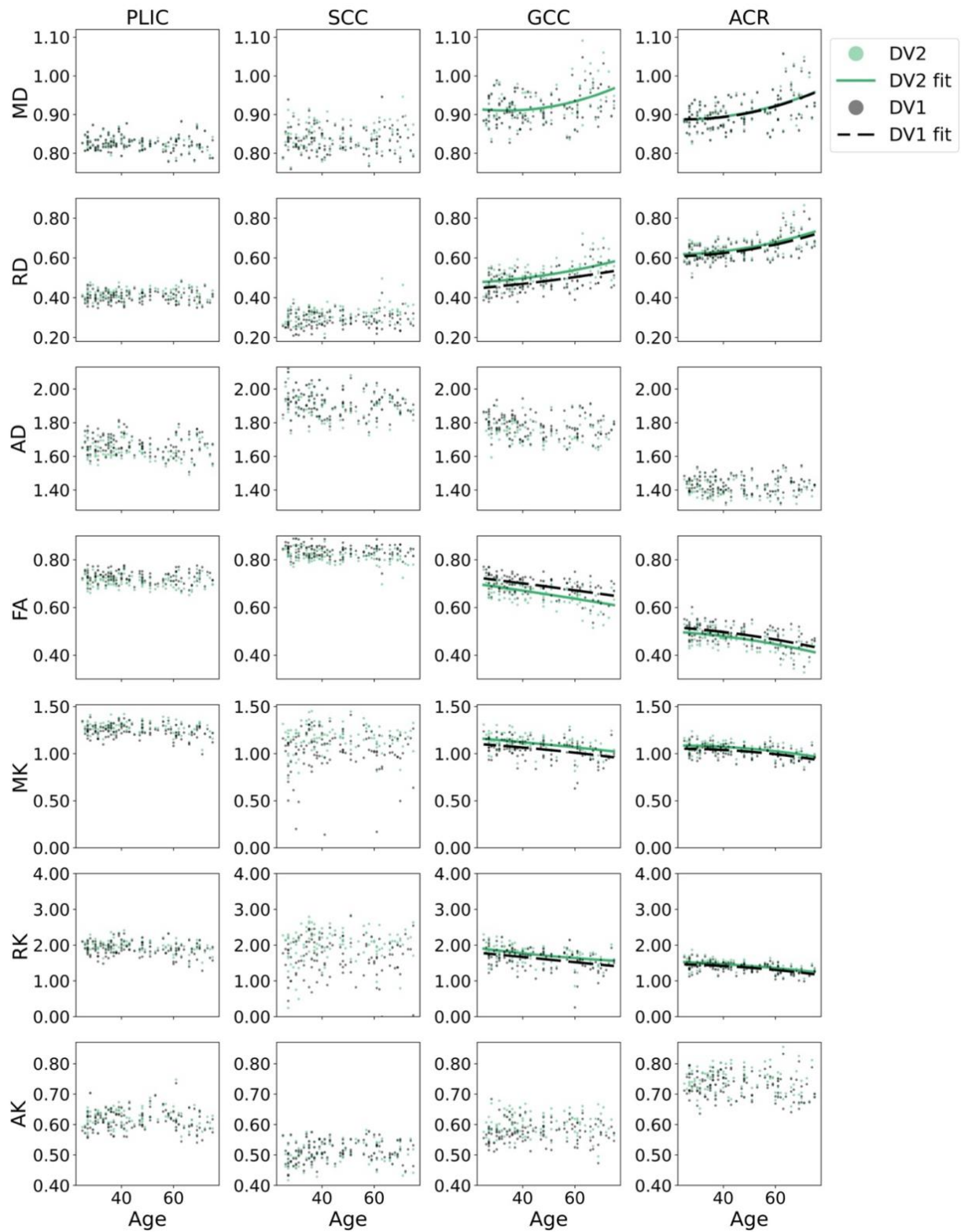

**Figure S8.** Age correlation with DTI and DKI parameters in white matter ROIs (median) from DESIGNER-Dv2 and DESIGNER-Dv1 pipeline of 120 healthy subjects (Prisma, TE = 95ms). Quadratic fits were plotted for statistically significant correlations with adjusted $R^2 > 0.1$ only.

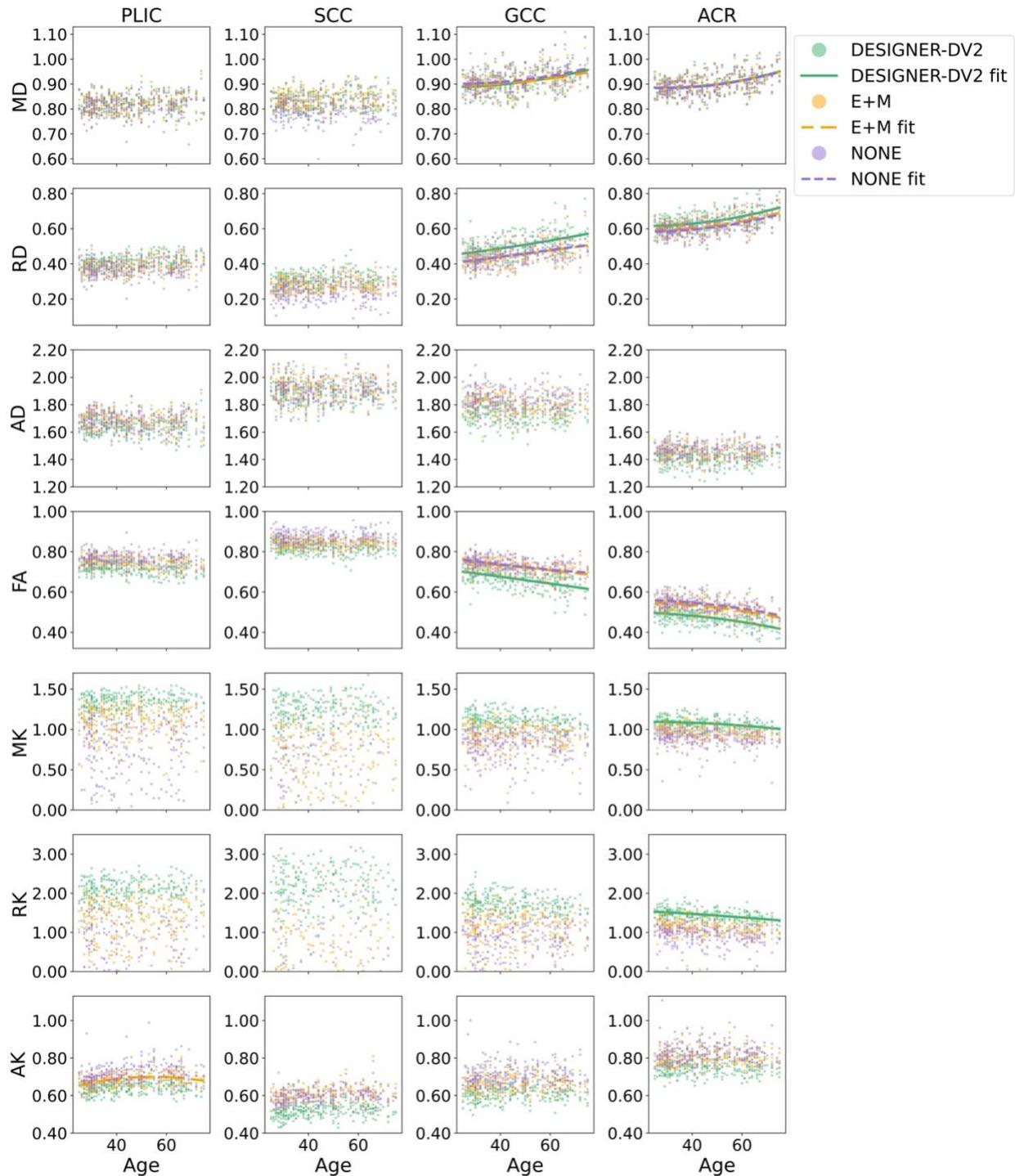

**Figure S9.** Age correlation with DTI and DKI parameters in white matter ROIs (median) from DESIGNER-Dv2 pipeline, E+M, or not preprocessing pipeline of 262 healthy subjects (Skyra, TE = 95ms). Quadratic fits were plotted for statistically significant correlations with adjusted $R^2 > 0.1$ only. MK and RK plots are zoomed in so not all datapoints are visible.

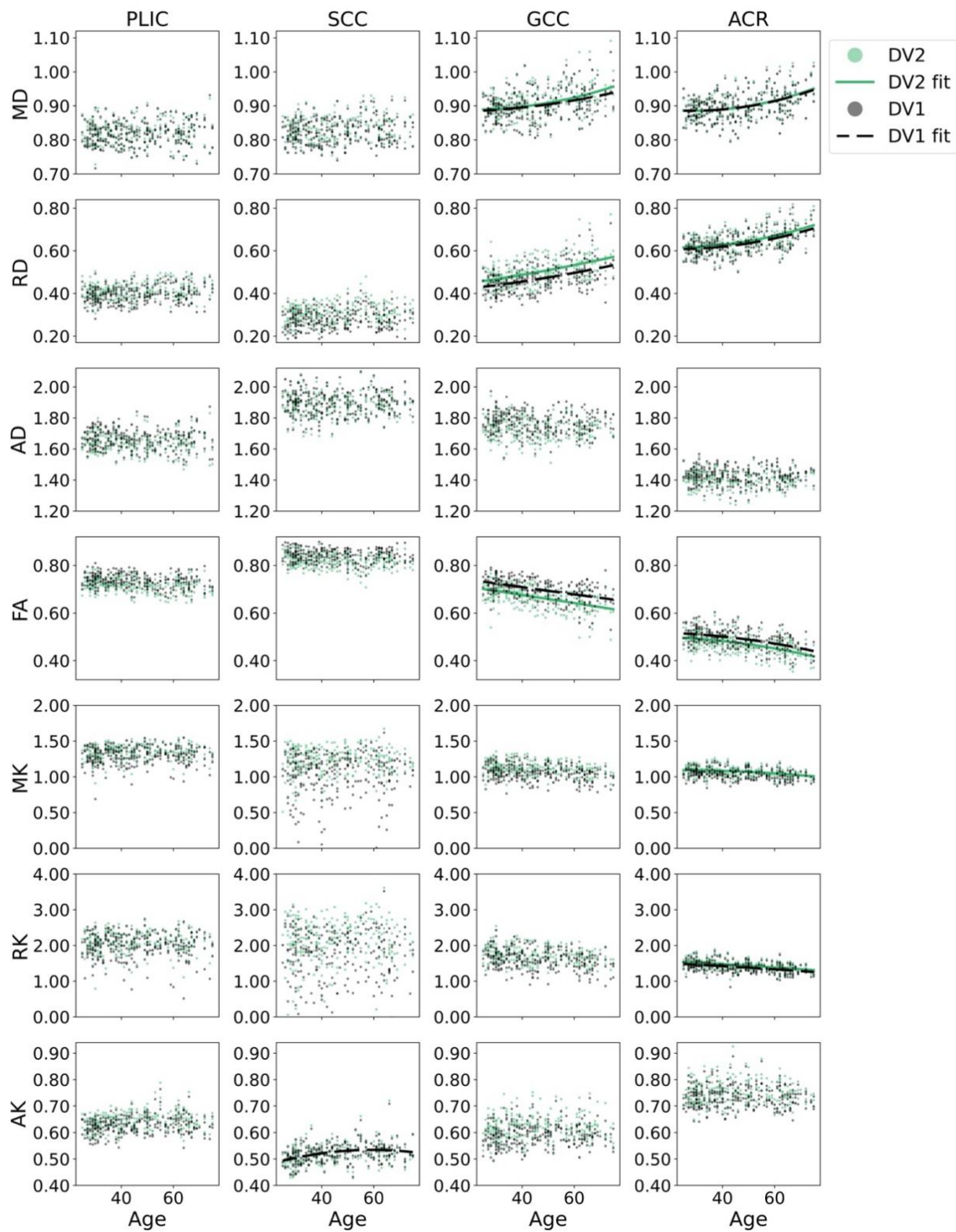

**Figure S10.** Age correlation with DTI and DKI parameters in white matter ROIs (median) from DESIGNER-Dv2 and DESIGNER-Dv1 pipeline of 262 healthy subjects (Skyra, TE = 95ms). Quadratic fits were plotted for statistically significant correlations with adjusted $R^2 > 0.1$ only.

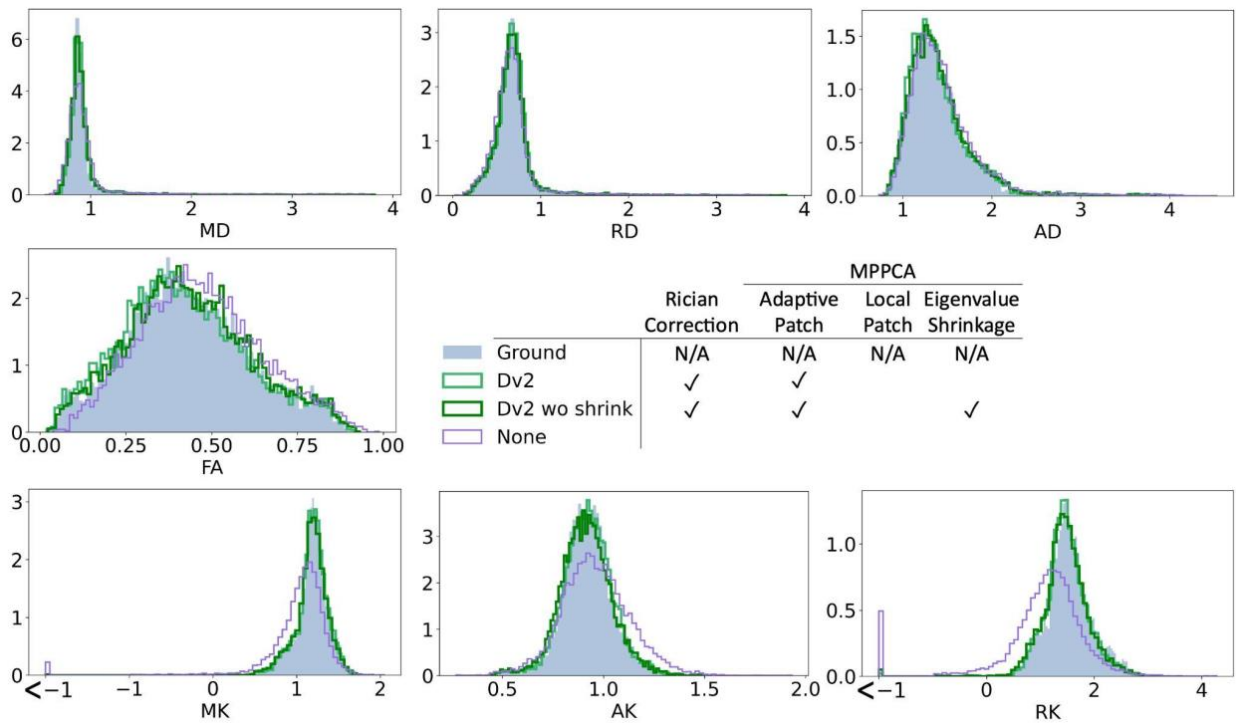

**Figure S11.** Probability density of DTI and DKI parameters in white matter of ground truth HCP phantom and HCP noise phantom (SNR 20) after adaptive patch denoising with eigenvalue shrinkage (with Rician bias correction) and adaptive patch denoising without eigenvalue shrinkage (with Rician bias correction).

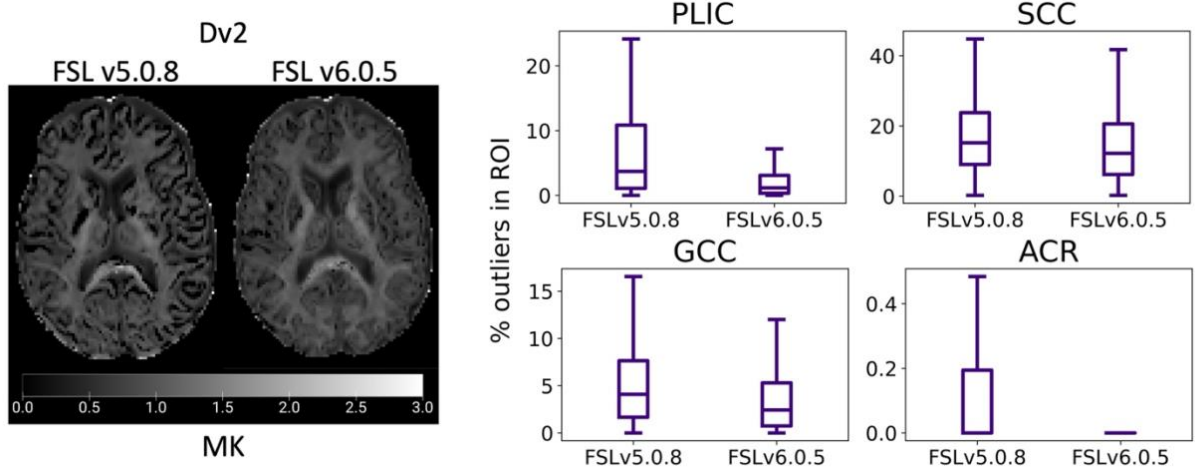

**Figure S12**. Left: MK maps from a healthy 69-year-old female after preprocessing with DESIGNER-Dv2 using different FSL versions (v5.0.8 and v6.0.5) for motion and eddy current correction. Right: Box plots of percent outliers (100*number of outliers in ROI/number of voxels in ROI) in each WM ROI of the dMRI parameter maps of 524 subjects based on DESIGNER-Dv2 pipeline using different FSL versions. One-way ANOVA showed percent outliers in each of the four ROIs were significantly different between FSL versions except in ACR.